\documentclass[onecolumn, ceqn, showpacs, showkeys, prd, onecolumn, byrevtex, nofootinbib, amsmath, amsfonts, amssymb]{revtex4-1}

\usepackage{amssymb,amsmath,mathtools,xcolor,graphicx,xspace,colortbl,ragged2e,rotating} %
\usepackage{textcomp}
\usepackage{amsfonts}  
\usepackage{amsmath}  
\usepackage{amssymb}  
\usepackage{boxedminipage}  
\usepackage{graphics}  
\usepackage{graphicx}  
\usepackage{ragged2e}  
\usepackage{tabulary}  
\usepackage{varioref}  
\usepackage{wrapfig}  
\usepackage{xcolor}  
\graphicspath{{FRACTIONAL2_3_graphics/}{FRACTIONAL2_3_tcache/}{FRACTIONAL2_3_gcache/}}
\DeclareGraphicsExtensions{.pdf,.eps,.ps,.png,.jpg,.jpeg}

\begin{document}\thinspace

\title{NEWTONIAN FRACTIONAL-DIMENSION GRAVITY AND MOND}
\author{Gabriele U. Varieschi
}
\email[E-mail me at: ]{gvarieschi@lmu.edu
}
\homepage[Visit: ]{http://gvarieschi.lmu.build
}
\affiliation{Loyola Marymount University, Los Angeles, CA 90045, USA
}

\date{
\today
}
\begin{abstract}
This paper introduces a
possible alternative model of gravity based on the theory of fractional-dimension spaces and its applications to Newtonian
gravity.  In particular, Gauss's law for gravity as well as other fundamental
classical laws are extended to a
$D$-dimensional metric space, where
$D$
can be a non-integer dimension.

We show a possible connection between this
Newtonian Fractional-Dimension Gravity (NFDG) and Modified Newtonian Dynamics (MOND), a leading alternative
gravity model which accounts for the observed properties of galaxies and other astrophysical
structures without requiring the dark matter hypothesis. The MOND acceleration constant
$a_{0} \simeq 1.2 \times 10^{ -10}\mbox{m}\thinspace \mbox{s}^{ -2}$
can be related to a natural scale length
$l_{0}$
in NFDG, i.e.,
$a_{0} \approx GM/l_{0}^{2}$, for astrophysical structures of mass
$M$, and the deep-MOND regime is present in regions of space where the dimension is reduced to
$D \approx 2$.

 For several fundamental
spherically-symmetric structures, we compare MOND results, such as the empirical Radial Acceleration
Relation (RAR), circular speed plots, and logarithmic plots of the observed radial acceleration
$g_{obs}$
vs. the baryonic radial acceleration
$g_{bar}$, with NFDG results. We show that our model is capable of reproducing these results using a variable local
dimension
$D\left (w\right )$, where
$w =r/l_{0}$
is a dimensionless radial coordinate. At the moment, we are unable to derive explicitly
this dimension function
$D\left (w\right )$
from first principles, but it can be obtained empirically in each case from the general
RAR.

 Additional work on the subject, including
studies of axially-symmetric structures, detailed galactic rotation curves fitting, and a possible
relativistic extension, will be needed to establish NFDG as a viable alternative model of gravity.
\end{abstract}
\maketitle

\section{\label{sect:intro}
Introduction
}
 This paper considers a possible generalization of the gravitational Gauss's law and
of the other standard laws of Newtonian gravity to lower dimensional cases, including fractional
(i.e., non-integer) dimensions. This analysis is based on the application of fractional calculus
(FC) \cite{MR0361633,MR1219954,MR1658022,Herrmann:2011zza,MR1890104} and fractional mechanics \cite{bookTarasov,bookZubair} to the classical laws of gravity. Fractional calculus is also commonly related to
fractal geometries, which might be relevant at galactic or cosmological scales in the universe \cite{Baryshev:2002tn,Nottale:2011zz,Calcagni:2016azd}. A generalized
``Newtonian Fractional-Dimension Gravity'' (NFDG) can be derived from the aforementioned principles.\protect\footnote{
We note that our NFDG is not a ``fractional'' theory in the sense used by other models in the literature \cite{Calcagni:2016azd}. The ``fractional'' Gauss's law that will be introduced in Sect. \ref{sect::gauss}, will be connected to a Weyl's fractional integral, but the resulting equations for the gravitational potential and field will be based on operators constructed with ordinary derivatives. Thus, our model has a fractal structure, due to the non-integer dimension of the metric space, and would be better described as ``Newtonian gravity in fractional dimensional spaces.'' However, we prefer to call it NFDG for simplicity's sake.
}

This analysis also shows a possible connection with Modified Newtonian Dynamics
(MOND) \cite{Milgrom:1983ca,Milgrom:1983pn,Milgrom:1983zz}, a leading alternative gravity model, originally introduced in 1983 as a possible
solution to the mass-discrepancy/dark matter (DM) problem, and later evolved into a relativistic
theory \cite{Bekenstein:2004ne,Sanders:2005vd}.
Recently, a strong correlation between the radial gravitational acceleration traced by galactic
rotation curves and that predicted by the observed distribution of baryons has been reported
\cite{McGaugh:2016leg,Lelli:2017vgz}, consistent with
MOND galactic dynamics, although this model still has weaknesses and issues in explaining the phenomenology of large-scale structures, galaxy clusters, cosmic microvawe background radiation, and others \cite{Famaey:2011kh}.

In this paper, we propose a possible explanation of this correlation based
on a ``variable-dimension'' effect in galactic structures, thus connecting MOND with
Newtonian fractional-dimension gravity.
It should be noted that our NFDG, as well as other similar models recently introduced (\cite{2020PhRvD.101l4029G}, see also a brief discussion in Sect. \ref{sect::appendix_multipole}), is not a fractional or a fractal version of MOND, but rather a MOND-like model reproducing the asymptotic behavior of MOND. In fact, MOND is a fully non-linear theory, while NFDG is inherently linear, in view of its fundamental equations which will be presented in Sect. \ref{sect::gauss}. However, in this work we also attempt to describe the transition between the two asymptotic regimes of MOND by assuming a continuous (slow) change in the variable dimension of the associated metric space.

In Sect. \ref{sect:MOND}, we describe in more
details the original MOND model and the related correlation between observed and baryonic galactic
rotational accelerations. In Sect. \ref{sect::gauss}, we
introduce the fractional generalization of Gauss's law and Newtonian gravity. In Sect. \ref{sect::galactic}, we show applications of these
methods to some basic spherical models and establish a connection with the MOND\ theory. Finally, in
Sect. \ref{sect::conclusion} conclusions are drawn
and possible future work on the subject is outlined.

\section{\label{sect:MOND}
MOND and galactic rotation curves
}
Modified Newtonian Dynamics was created in 1982
by Milgrom and first published in 1983 \cite{Milgrom:1983ca,Milgrom:1983pn,Milgrom:1983zz} as a possible solution to
the mass discrepancy problem in galaxies, galaxy systems, and other stellar systems. The
original proposal called for a modification of Newtonian dynamics in terms of inertia or gravity, in
order to describe the motion of bodies in the gravitational field of a galaxy, or a cluster of
galaxies, without the need of any hidden mass, i.e., without any DM contribution. At the core of the
MOND\ model is an acceleration constant
$a_{0}$, whose currently estimated value is \cite{McGaugh:2016leg,Lelli:2017vgz}:

\begin{equation}a_{0} \equiv g_{\dag } =1.20 \pm 0.02\ \text{(random)} \pm 0.24\ \text{(syst)} \times 10^{ -10}\ \mbox{}\ \mbox{m}\thinspace \mbox{s}^{ -2} \label{eq2.1}
\end{equation}
and which represents the acceleration scale below which MOND corrections are
applied.\protect\footnote{ In the latest papers in the literature (such as  \cite{McGaugh:2016leg,Lelli:2017vgz}), this acceleration scale constant is indicated as
$g_{\dag }$, when derived by fitting galactic data with an empirical law. See Eq. (\ref{eq2.6}) later in this section.
}

The modification to Newtonian dynamics can be expressed in two possible forms
\cite{Bekenstein:1984tv}:

\begin{gather}m\mu (a/a_{0})\mathbf{a} =\mathbf{F} \label{eq2.2} \\
\mu (g/a_{0})\mathbf{g} =\mathbf{g}_{N} , \nonumber \end{gather}
where the former indicates a modification to the law of inertia (Newton's second law):
$\mathbf{F}$
is an arbitrary static force and
$m$
is the (gravitational) mass of the accelerated test particle. For the force of gravity,
$\mathbf{F} =m\mathbf{g}_{N}$, where
$\mathbf{g}_{N} = - \nabla \phi _{N}$
and
$\phi _{N}$
is the usual Newtonian gravitational potential derived from the standard Poisson
equation. Therefore, this first form of modified dynamics applies to any type of force and changes
the law of inertia, since the acceleration
$\mathbf{a}$
is replaced by
$\mu (a/a_{0})\mathbf{a}$.

 On the contrary, the second line in Eq. (\ref{eq2.2}) modifies just the gravitational field
$\mathbf{g}$, obtained from the Newtonian
$\mathbf{g}_{N}$, while leaving the law of inertia ($m\mathbf{a} =\mathbf{F}$) unchanged. The two formulations are practically equivalent, but conceptually
different:\ the former modifies Newton's laws of motion, while
the latter modifies Newton's law of universal gravitation. In both cases, the modification follows
from the \textit{interpolation function }$\mu (x) \equiv \mu (a/a_{0})\text{}$
or
$\mu \text{}(x) \equiv \mu (g/a_{0})$, respectively.

MOND postulates that:

\begin{equation}\mu (x) \approx \left \{\begin{array}{c}1\text{ for}\ \text{}x \gg 1\ \text{(Newtonian regime),  }\text{}\text{
} \\
x\ \text{for}\ x \ll 1\ \text{(deep-MOND regime).}\text{
}\end{array}\right . \label{eq2.3}
\end{equation}

 Originally, Milgrom used simple forms for the interpolation function, such as the
``standard'' form
$\mu _{2}(x) =x/\sqrt{1 +x^{2}}$
or
$\mu (x) =1 -e^{ -x}$\cite{Milgrom:1983pn,Milgrom:1984ij}, while recently other
forms have become more popular, such as the ``simple'' interpolation function
$\mu _{1}(x) =x/(1 +x)$, or the general family of functions
$\mu _{n}(x) =x(1 +x^{n})^{ -1/n}$, of which
$\mu _{1}$
and
$\mu _{2}$
are special cases. Also, following the second line in Eq. (\ref{eq2.2}), it has become customary \cite{McGaugh:2008nc} to
invert this relation into the following:

\begin{equation}\mathbf{g} =\mu ^{ -1}(x)\mathbf{g}_{N} \equiv \nu (y)\mathbf{g}_{N}\ \text{with}\ y =g_{N}/a_{0} . \label{eq2.4}
\end{equation}

Several
$\nu (y)$
functions were introduced in the literature over the years (see Ref. \cite{McGaugh:2008nc} for full details), but
we will consider in the following two main families:

\begin{gather}\nu _{n}(y) =\left (\frac{1}{2} +\frac{1}{2}\sqrt{1 +4y^{ -n}}\right )^{1/n} , \label{eq2.5} \\
\widehat{\nu }_{n}(y) =\left [1 -\exp \left ( -y^{n/2}\right )\right ]^{ -1/n} . \nonumber \end{gather}
The first family in the last equation is the inverse of the family
$\mu _{n}(x)$
described above, while the second one corresponds to interpolation functions which are
similar to the ``simple'' function on galaxy scales ($ \sim a_{0}$) while having no impact in the inner solar system ($ \sim 10^{8}a_{0}$) \cite{McGaugh:2008nc}.
Due to the exponential function in their definition, the functions
$\widehat{\nu }_{n}(y)$
cannot be easily inverted into corresponding
$\widehat{\mu }_{n}(x)$
functions, but the particular choice
$\widehat{\nu }_{1}(y) =\left [1 -\exp \left ( -y^{1/2}\right )\right ]^{ -1}$
has recently become the favorite interpolation function \cite{McGaugh:2016leg,Lelli:2017vgz}.

 Continuing our brief historical outline of MOND (see Refs. \cite{Milgrom:2001ny,Bekenstein:2007iq,Famaey:2011kh} for general
reviews), in order to ensure standard conservation laws Milgrom and Bekenstein introduced an
Aquadratic Lagrangian theory (AQUAL) \cite{Bekenstein:1984tv}, in the context of MOND as modified gravity. This was based on a Lagrangian
function which generalized the Newtonian one and on a modified Poisson equation,
$ \nabla  \cdot \left [\mu \left (\left \vert  \nabla \phi \right \vert /a_{0}\right ) \nabla \phi \right ] =4\pi G\rho $, from which the original MOND relation in the second line of Eq. (\ref{eq2.2}) would follow in cases of high symmetry (spherical, cylindrical, or
plane).

This version of MOND was then interpreted by Milgrom \cite{Milgrom:1992hr,Milgrom:1998sy} as an effective theory due to vacuum effects of
cosmological origin. This view was also supported by the numerical coincidences:
$a_{0} \approx cH_{0}$
and
$a_{0} \approx c^{2}\Lambda ^{1/2}$, linking the MOND acceleration
$a_{0}$
with the Hubble constant
$H_{0}$
and the cosmological constant
$\Lambda $. The AQUAL model then evolved into different relativistic versions, such as RAQUAL and
phase-coupled gravity--PCG \cite{Bekenstein:2007iq}, which were
eventually replaced by other relativistic MOND theories (Tensor-Vector-Scalar theories \cite{Bekenstein:2004ne,Sanders:2005vd}).

Other relativistic models of gravity \cite{Mannheim:2005bfa,Vagnozzi:2017ilo} also reproduce MOND phenomenology at low energies. Since our analysis will
be confined to non-relativistic effects, there is no need to introduce further elements of
relativistic MOND, but we will mention Verlinde's emergent gravity (EG) framework \cite{2017ScPP....2...16V}, an attempt to explain gravity as an entropic force (see also a brief discussion in Sect. \ref{sect::appendix_multipole}), and the recent proposal by Skordis et al. \cite{Skordis:2020eui}.

 Back to the non-relativistic model, the main successes of MOND are well-known. In
the spirit of ``Keplerian''\ laws, they can be summarized as
the three laws of rotationally-supported galaxies (from Ref. \cite{McGaugh:2014xfa}):

\begin{enumerate}
\item Rotation curves attain an approximately constant velocity (asymptotic or
flat rotation velocity
$V_{f}$) that persists indefinitely (flat rotation curves).

\item The observed baryonic mass scales as the fourth power of the amplitude of
the flat rotation curve (the ``baryonic'' Tully-Fisher relation-BTFR:
$M_{bar} \sim V_{f}^{4}$).

\item There is a one-to-one correspondence between the radial force and the
observed distribution of baryonic matter (the mass discrepancy-acceleration relation
$M_{tot}/M_{bar} \simeq V_{obs}^{2}/V_{bar}^{2}$).
\end{enumerate}

The first point is the main original MOND\ prediction from Eqs. (\ref{eq2.2})
and (\ref{eq2.3}), regardless of the chosen interpolation
function \cite{Milgrom:1983pn}: at large galactic radii (deep-MOND
regime) we can replace
$g_{N} \approx GM/r^{2}$
and
$a =g =V^{2}/r$
in those two equations and obtain
$V^{4}(r) \equiv V_{f}^{4} \approx GMa_{0}$, from which
$V_{f} \approx \sqrt[{4}]{GMa_{0}}$, where
$M$
is the total mass of the galaxy. This initial consideration then evolved into detailed
fitting of galactic rotation curves of different shapes without using dark matter (see
Refs. \cite{Bekenstein:2007iq,McGaugh:2008nc,McGaugh:2014xfa} for some examples) and firmly established MOND as an alternative to the DM
hypothesis.

 The second point follows directly from the previous arguments, in particular from
$V_{f}^{4} \approx GMa_{0}$, and from the observed strong correlation (Tully-Fisher relation \cite{Tully:1977fu,Lelli:2015wst,Lelli:2019igz}) between galactic total
luminosity
$L$
and typical rotational velocity
$V_{f}$:
$L\text{} \propto V_{f}^{\alpha }$
with
$\alpha  \sim 4$, and by assuming also that the total galactic luminosity is simply proportional to the
total mass
$M$. This connection between light and matter is also manifest in more subtle ways:\ features in galactic luminosity profiles usually correspond to
features in the rotation curves and vice-versa (the so-called \textit{Renzo's rule
}\cite{Sancisi:2003xt}).

 The third point has recently evolved into a very precise statement relating the
radial acceleration
$g_{obs}$
traced by rotation curves with the radial acceleration
$g_{bar}$
predicted by the observed distribution of baryonic matter \cite{McGaugh:2016leg,Lelli:2017vgz}. Using astrophysical data from the Spitzer Photometry and Accurate Rotation Curves
(SPARC) database \cite{Lelli:2016zqa}, a sample of 175 galaxies with
new photometry and high-quality rotation curves, McGaugh and collaborators were able to produce an
empirical fit to all the data points (radial acceleration relation - RAR) as follows:

\begin{equation}g_{obs} =\frac{g_{bar}}{1 -e^{ -\sqrt{g_{bar}/g_{\dag }}}} , \label{eq2.6}
\end{equation}
where
$g_{\dag }$
is an empirical acceleration parameter, corresponding to the MOND (theoretical)
acceleration scale
$a_{0}$, whose value we already reported in Eq. (\ref{eq2.1})
above. While Eq. (\ref{eq2.6}) represents a pure empirical fit,
in spirit similar to Kepler's third law, it can be noted immediately that it corresponds to one of
the MOND interpolation functions described above, namely, the particular function
$\widehat{\nu }_{1}(y) =\left [1 -\exp \left ( -y^{1/2}\right )\right ]^{ -1}$, once we identify$\ y =g_{N}/a_{0} \equiv g_{bar}/g_{\dag }$
and
$\widehat{\nu }_{1} =g/g_{N} \equiv g_{obs}/g_{bar}$.

 It should be also noted that the above fit involved 2693 data points from 153
rotationally supported (spiral and irregular) galaxies \cite{McGaugh:2016leg}, spanning over a large range of physical properties, such as rotation
velocities, luminosities, effective surface brightness, etc. The baryonic acceleration
$g_{bar}$
was computed first by solving numerically the standard Poisson equation
$ \nabla ^{2}\phi _{bar} =4\pi G\varrho _{bar}$, with the baryonic mass density
$\rho _{bar}$
obtained from near-infrared ($3.6\ \mbox{{\textmu}m}$) data tracing the stellar mass distribution and from
$21$$\mbox{cm}$
hydrogen line data tracing the atomic gas, and then by deriving
$g_{bar}$
from the gravitational potential:
$g_{bar} =\genfrac{\vert }{\vert }{}{}{ \partial \phi _{bar}}{ \partial R}$. The observed acceleration
$g_{obs}$
was obtained directly from the rotation curves as
$g_{obs} =V^{2}/R$. By using a single value for the stellar mass-to-light ratio
$\Upsilon _{ \ast }^{[3.6]} =0.50M_{ \odot }/L_{ \odot }$\cite{McGaugh:2016leg}, the fit in Eq. (\ref{eq2.6}) provides a universal empirical equation based just on the MOND
parameter
$g_{\dag } \equiv a_{0}$: in Fig. 3 of Ref.
\cite{McGaugh:2016leg}
nearly 2700 data points are fitted by the RAR and individual galaxies are indistinguishable in that
analysis. In Sects. \ref{sect::gauss} and \ref{sect::galactic} we will provide a possible explanation of
this fundamental empirical relation using NFDG.

 In more recent work  \cite{Lelli:2017vgz}, from the original SPARC\ database of
late-type-galaxies (spiral and irregular), the galaxy sample was extended to include
early-type-galaxies (elliptical and lenticular) and dwarf spheroidal galaxies, confirming the RAR
and, as a consequence, also other dynamical properties of galaxies, like the Tully-Fisher and
Faber-Jackson relations, the ``baryon-halo'' conspiracies, and Renzo's rule. Also, in Ref.  \cite{Li:2018tdo} the 175 galaxies in the SPARC database were checked
individually against the RAR, allowing for galaxy-to-galaxy variations of the acceleration scale
$g_{\dag }$. The result of this analysis favored a single value of
$g_{\dag }$, consistent with the action of a single effective force law.

Several other experimental observations are explained by MOND and related
relativistic versions (see \cite{Famaey:2011kh} for a complete
discussion), but it should be noted that MOND and its generalizations do not adequately explain
properties of galaxy clusters, globular clusters, etc., and are not particularly suited to form the
basis of a cosmological model \cite{Famaey:2011kh}. In particular, MOND cannot completely eliminate the need for dark matter in all astrophysical systems: galaxy clusters show a residual mass discrepancy when analyzed using MOND, globular clusters and their velocity dispersion profiles are not perfectly fitted by MOND, and cosmic microwave background measurements place constraints on alternative gravity theories, such as MOND and EG \cite{2020arXiv200700555P}. With respect to these issues and as a standard cosmological model, the
$\Lambda CDM$
concordance model \cite{Ostriker:1995su} is still the
favorite, most successful, cosmological theory.

\section{\label{sect::gauss}
Fractional Gauss's law and Newtonian gravity
}
An interesting pedagogical problem is the
analysis of classical theories, such as Maxwell's electrodynamics or Newtonian gravity, in a
lower-dimensional space-time. Rather than using standard
$3 +1$
space-time, we can decrease the spatial dimension and consider these classical theories in
$2 +1$, or even
$1 +1$
space-times. Standard textbooks do not usually discuss these cases, and very few papers
(see \cite{Boito:2018rdh,McDonald:2018} and references therein) are
available in the literature. The study of the laws of physics for cases of dimension
$D \neq 3$
was also used at times as a ``proof'' of the tri-dimensionality of space (see the original
analysis by Ehrenfest in 1920 \cite{doi:10.1002/andp.19203660503} and
the more philosophical discussion in Ref. \cite{2005SHPMP..36..113C}).

 Typically, only Maxwell's electrodynamics in lower-dimensional spaces is discussed
in the literature \cite{Boito:2018rdh,McDonald:2018}, but results for electrostatics can be easily adapted to
Newtonian gravity. The main issue is whether Coulomb's law should be altered in a lower-dimensional
space situation, or remain the usual inverse-square law. This crucial point was clarified by
Lapidus \cite{1981AmJPh..49..807L,1982AmJPh..50..155L},
who argued that
Gauss's law in a two-dimensional space implies an inverse-linear Coulomb's law and
thus a
logarithmic potential, while in a one-dimensional space the electric field is constant in
magnitude and the potential linear. These electrostatic potentials were then used to study one and
two-dimensional hydrogen atoms \cite{1985AmJPh..53..893A,Castro:2010yb}.

 In the following, we will start from the
analysis of electrostatics in two spatial dimensions outlined in Ref. \cite{Boito:2018rdh} and in the unpublished Ref. \cite{McDonald:2018}, adapt it to the gravitational case and then generalize it to an arbitrary
fractional dimension
$D$. The laws of electrostatics are easily converted into equivalent gravitational laws by
replacing
$1/\epsilon _{0}\text{}$
with
$ -4\pi G$, where
$\epsilon _{0}\text{ is the permittivity of free space and}\text{ }G\text{ is}\text{}\text{}\text{}$
the standard gravitational constant.\protect\footnote{ SI\ units will be used throughout this
paper, unless otherwise noted.
} In this way, standard Gauss's law for the electric field
$\mathbf{E}\text{}$
of a point charge
$q$, placed at the origin in a three-dimensional ($D =3$) space, describes the (outward) flux of
$\mathbf{E}$
through a spherical surface
$S$
of radius
$r$
as:
$\Phi _{E}^{(3)} =\left \vert \mathbf{E}\right \vert r^{2}\int _{S}d\Omega  =\left \vert \mathbf{E}\right \vert 4\pi r^{2} =q/\epsilon _{0}$.

The equivalent law for the (attractive)
gravitational field
$\mathbf{g}$
due to a point mass
$m$
is:

\begin{equation}\Phi _{g}^{(3)} = -\left \vert \mathbf{g}\right \vert r^{2}\int _{S}d\Omega  = -\left \vert \mathbf{g}\right \vert 4\pi r^{2} = -4\pi Gm , \label{eq3.1}
\end{equation}
which yields the usual inverse-square law for the gravitational field:
$\left \vert \mathbf{g}\right \vert  =Gm/r^{2}$.

Gauss's law can be generalized to any
(positive) dimension
$D$
by considering a hypersphere
$S$
of radius
$r$
as a Gaussian surface with
$\int _{S}d\Omega _{D} =\frac{2\pi ^{D/2}}{\Gamma (D/2)}$, which follows from standard dimensional regularization techniques commonly used in
quantum field theory (see, for example, \cite{1995iqft.book.....P} page
249, or the original Refs. \cite{Bollini:1972ui,tHooft:1972tcz,Wilson:1972cf}). Similar regularization
methods on fractal
space-times \cite{1987JPhA...20.3861S,bookTarasov,bookZubair} generalize the integral of a
spherically-symmetric function
$f =f(r)$
over a fractal
$D$-dimensional metric space
$W$
as follows:

\begin{equation}\int _{W}fd\mu _{H} =\frac{2\pi ^{D/2}}{\Gamma (D/2)}\int _{0}^{\infty }f(r)r^{D -1}dr , \label{eq3.2}
\end{equation}
where
$\mu _{H}$
denotes an appropriate Hausdorff measure over the space.

In fractional calculus,\protect\footnote{ For a basic introduction to FC and some elementary physics application see
Ref. \cite{Varieschi:2018}.
} the integral on the right-hand side of Eq. (\ref{eq3.2}) is related to the Weyl's fractional integral\protect\footnote{ In FC there are several possible definitions of fractional integrals and
derivatives and notation also differs in leading textbooks \cite{MR0361633,MR1219954,MR1658022,Herrmann:2011zza,MR1890104}.  For example,
following Ref. \cite{MR1219954}, the Riemann integral is defined as
$\,_{c}D_{x}^{ -\nu }f(x) =\frac{1}{\Gamma (\nu )}\int _{c}^{x}(x -t)^{\nu  -1}f(t)dt$, the Liouville integral ($c = -\infty $) as
$\,_{ -\infty }D_{x}^{ -\nu }f(x) =\frac{1}{\Gamma (\nu )}\int _{ -\infty }^{x}(x -t)^{\nu  -1}f(t)dt$, the Riemann-Liouville ($c =0$) as
$\,_{0}D_{x}^{ -\nu }f(x) =\frac{1}{\Gamma (\nu )}\int _{0}^{x}(x -t)^{\nu  -1}f(t)dt$, and the Weyl fractional integral as
$\,_{x}W_{\infty }^{ -\nu }f(x) =\frac{1}{\Gamma (\nu )}\int _{x}^{\infty }(t -x)^{\nu  -1}f(t)dt$. Some textbooks \cite{bookTarasov} prefer to distinguish
between left-sided integrals, such as
$\,_{c}D_{x}^{ -\nu }f(x) =\frac{1}{\Gamma (\nu )}\int _{c}^{x}(x -t)^{\nu  -1}f(t)dt$, and right-sided integrals, such as
$\,_{x}D_{c}^{ -\nu }f(x) =\frac{1}{\Gamma (\nu )}\int _{x}^{c}(t -x)^{\nu  -1}f(t)dt$; in this way the Weyl integral is the right-sided version of the left-sided Liouville
integral above. 
} defined as
$W^{ -D}f(x) =\frac{1}{\Gamma (D)}\int _{x}^{\infty }(t -x)^{D -1}f(t)dt$, so that Eq. (\ref{eq3.2}) can also be written as
\cite{1987JPhA...20.3861S,bookTarasov}:

\begin{equation}\int _{W}fd\mu _{H} =\frac{2\pi ^{D/2}\Gamma (D)}{\Gamma (D/2)}W^{ -D}f(0). \label{eq3.3}
\end{equation}
This equation connects the integral over a fractal space
$W$
of dimension
$D$
with an integral of fractional order up to the numerical factor
$2\pi ^{D/2}\Gamma (D)/\Gamma (D/2)$, thus establishing a direct relation between fractal space-time and fractional calculus.

 In view of this discussion, the generalized
gravitational Gauss's law becomes:
\begin{equation}\Phi _{g}^{(D)} = -\left \vert \mathbf{g}\right \vert r^{D -1}\int _{S}d\Omega _{D} = -\left \vert \mathbf{g}\right \vert r^{D -1}\frac{2\pi ^{D/2}}{\Gamma (D/2)} = -4\pi Gm_{(D)} , \label{eq3.4}
\end{equation}
where
$m_{(D)}$
represents a ``point-mass'' in a fractal
$D$-dimensional space. Since
$\Gamma (3/2) =\frac{\sqrt{\pi }}{2}$, for
$D =3$
the last equation reduces to the standard law
$\left \vert \mathbf{g}\right \vert  =\frac{Gm_{(3)}}{r^{2}}$, with
$m_{(3)} =m$
representing the standard mass measured in kilograms. For
$D =2$
and
$D =1$, we obtain respectively
$\left \vert \mathbf{g}\right \vert  =\frac{2Gm_{(2)}}{r}$
and
$\left \vert \mathbf{g}\right \vert  =2\pi Gm_{(1)}$
(since
$\Gamma (1) =0 ! =1$
and
$\Gamma (1/2) =\sqrt{\pi }$
). Assuming that
$\left \vert \mathbf{g}\right \vert $ (in $\mbox{m}\thinspace \mbox{s}^{ -2}$) and
$G =6.674 \times 10^{ -11}$$\mbox{m}^{3}\thinspace \mbox{kg}^{ -1}\thinspace \mbox{s}^{ -2}$
retain the same physical dimensions (and the same value for
$G$) in any
$D$-dimensional space, it is easy to check that
$m_{(2)}$
and
$m_{(1)}$
will have dimensions of mass per unit length ($\mbox{kg}\thinspace \mbox{m}^{ -1}$) $\mbox{}$and mass per unit surface ($\mbox{kg}\thinspace \mbox{m}^{ -2}$), respectively.

 This is consistent with the general idea \cite{McDonald:2018} that electrostatics (or Newtonian gravity) in two spatial dimensions ($x ,y$) should be equivalent to that in three spatial dimensions ($x ,y ,z$) for situations in which the
$3$-dimensional charge (mass) distributions are independent of
$z$, i.e., lines of constant charge (mass) density in the
$z$
direction, whose fields scale like the inverse of the distance. Similarly, in one spatial
dimension ($x$), we should consider as sources surfaces of constant charge (mass) density independent of
($y ,z$), i.e., surfaces parallel to the ($y ,z$) plane, whose fields are uniform.

 From Eq. (\ref{eq3.4}), we can obtain the
general expression for the gravitational field as
$\left \vert \mathbf{g}\right \vert  =2\pi ^{1 -D/2}\Gamma (D/2)\frac{Gm_{(D)}}{r^{D -1}}$, with
$m_{(D)}$
measured in
$\mbox{kg}\thinspace $$\mbox{m}^{D -3}$. To avoid this awkward unit for masses in
$D$-dimensional spaces, we prefer to redefine the mass as
$m_{(D)} =\widetilde{m}_{(D)}/l_{0}^{3 -D}$, where
$l_{0}$
represents a scale length (in meters) and
$\widetilde{m}_{(D)}$
is now measured in kilograms. Our preferred expression for the gravitational field becomes:

\begin{equation}\left \vert \mathbf{g}\right \vert  =2\pi ^{1 -D/2}\Gamma (D/2)\frac{G\widetilde{m}_{(D)}}{l_{0}^{2}}\frac{1}{(r/l_{0})^{D -1}} \approx \frac{Gm_{\left (3\right )}}{l_{0}^{2}}\frac{1}{(r/l_{0})^{D -1}} . \label{eq3.5}
\end{equation}

In the last equation, we have also assumed that
$m \equiv m_{(3)} \approx 2\pi ^{1 -D/2}\Gamma (D/2)\widetilde{m}_{(D)}\text{ for}$ simplicity's sake. This is consistent with the discussion above that a
$3$-dimensional mass
$m_{(3)}$
should be replaced by
$2\widetilde{m}_{(2)} =2m_{(2)}l_{0}$
in 2-dim, and by
$2\pi \widetilde{m}_{(1)} =2\pi m_{(1)}l_{0}^{2}$
in 1-dim, where
$m_{(2)}$
and
$m_{(1)}$
are linear and surface mass densities, respectively. The scale length
$l_{0}$
in Eq. (\ref{eq3.5}) is still undetermined, but the
combination
$Gm_{(3)}/l_{0}^{2}$
represents a scale acceleration for the field of a point particle
$m =m_{(3)}$
in a fractal space of dimension
$D$. It is possible to identify tentatively this scale acceleration with the similar MOND
acceleration parameter
$a_{0}$:\protect\footnote{ In the literature \cite{Sanders:2005vd}, the
quantity
$r_{m} =\sqrt{GM/a_{0}}$
is sometimes called the MOND radius. In view of Eq. (\ref{eq3.6}), our fundamental scale length
$l_{0}$
is equivalent to the MOND radius
$r_{m}$.
}

\begin{equation}a_{0} \approx \frac{Gm_{\left (3\right )}}{l_{0}^{2}} , \label{eq3.6}
\end{equation}
and consider the deep-MOND regime as being equivalent to the
$D =2$
case of Eq. (\ref{eq3.5}),

\begin{equation}\left \vert \mathbf{g}\right \vert _{deep -MOND} \approx \frac{Gm_{\left (3\right )}}{l_{0}^{2}}\frac{1}{(r/l_{0})} =\frac{a_{0}l_{0}}{r} . \label{eq3.7}
\end{equation}

 This approach is consistent with the MOND analysis of circular motion of a test
object in the field generated by a total mass
$M$. As seen in Sect. \ref{sect:MOND}, MOND\ computes the flat (asymptotic) rotational speed
$V_{f}$
as
$V_{f}^{4} =GMa_{0}$, in strong agreement with the baryonic Tully-Fisher relation. Combining our last two
equations with
$m_{\left (3\right )} =M$
and setting
$\left \vert \mathbf{g}\right \vert _{deep -MOND} =V_{f}^{2}/r$, as in standard Newtonian dynamics, we obtain the same result
$V_{f}^{4} =GMa_{0}$.

Furthermore, our approach is also consistent with the (simplified) MOND analysis of
binary galaxies, considered as point-like masses in circular orbits \cite{Milgrom:1983zz}. Assuming that the binaries contain galaxies of equal mass
$M$, denoting with
$R$
the true separation and with
$V$
the velocity difference, standard Newtonian gravity computes
$V^{2} =2GM/R$, while MOND  \cite{Milgrom:1983zz} obtains instead
$V^{4} =4a_{0}GM$. Since the reduced mass of the system is
$M/2$, and using Eqs. (\ref{eq3.6}) and (\ref{eq3.7}) with
$m_{\left (3\right )} =M$, we obtain
$\frac{M}{2}\frac{V^{2}}{R} =M\left \vert \mathbf{g}\right \vert _{deep -MOND} =\frac{Ma_{0}l_{0}}{R}$, from which
$V^{2} =2a_{0}l_{0} =2a_{0}\sqrt{GM/a_{0}}$
and, by squaring, we obtain the same MOND relation for
$V^{4}$
as above.

The two simple cases just described also confirm our choice to set
$m \equiv m_{(3)} \approx 2\pi ^{1 -D/2}\Gamma (D/2)\widetilde{m}_{(D)}\text{}$
for any value of
$D$. For example, a unit mass
$m =m_{(3)} =1\ \mbox{kg}$
in a three-dimensional space would effectively correspond to
$\widetilde{m}_{(2)} =\frac{1}{2}\mbox{kg}$
in a two-dimensional space, or to
$\widetilde{m}_{(1)} =\frac{1}{2\pi }\mbox{kg}$
in a one-dimensional space.\protect\footnote{ More generally, it should be noted that we could have defined
$a_{0} \approx C^{2}\frac{Gm_{\left (3\right )}}{l_{0}^{2}}$,
$l_{0} \approx C\sqrt{\frac{Gm_{\left (3\right )}}{a_{0}}}$
with
$C >0$
constant. This choice would also yield
$V_{f}^{4} =GMa_{0}$
for the flat rotational speed and
$V^{4} =4a_{0}GM$
for a simple binary system of equal mass
$M$, using our equations with
$D =2$. We have opted to simply set
$C =1$
in Eq. (\ref{eq3.6}), which also justifies the assumption in the right-hand side of Eq. (\ref{eq3.5}).
}

 Considering Eq. (\ref{eq3.5}) for an
infinitesimal source mass
$d\widetilde{m}_{\left (D\right )}$
and integrating over the D-dimensional source volume
$V_{D}$:

\begin{equation}\mathbf{g}(\mathbf{w}) = -\frac{2\pi ^{1 -D/2}\Gamma (D/2)G}{l_{0}^{2}}{\displaystyle\int _{V_{D}}}d\widetilde{m}_{(D)}\frac{\left (\mathbf{x} -\mathbf{x}^{ \prime }\right )/l_{0}}{(\left \vert \mathbf{x} -\mathbf{x}^{ \prime }\right .\vert /l_{0})^{D}} = -\frac{2\pi ^{1 -D/2}\Gamma (D/2)G}{l_{0}^{2}}{\displaystyle\int _{V_{D}}}\widetilde{\rho }(\mathbf{w}^{ \prime })\frac{\mathbf{w} -\mathbf{w}^{ \prime }}{\left \vert \mathbf{w} -\mathbf{w}^{ \prime }\right .\vert ^{D}}d^{D}\mathbf{w}^{ \prime } , \label{eq3.8}
\end{equation}
where the source mass
$d\widetilde{m}_{(D)}$
is described by the position
$\mathbf{x}^{ \prime }$. In this equation, we have also introduced dimensionless coordinates
$\mathbf{w} \equiv \mathbf{x}/l_{0}$
for the field point and
$\mathbf{w}^{ \prime } \equiv \mathbf{x}^{ \prime }/l_{0}$
for the source point. It should be noted that the mass ``density''
$\widetilde{\rho }\left (\mathbf{w}^{ \prime }\right )$
is actually measured in kilograms, since
$d\widetilde{m}_{\left (D\right )} =\widetilde{\rho }\left (\mathbf{w}^{ \prime }\right )d^{D}\mathbf{w}^{ \prime }$
(or
$\widetilde{\rho }\left (\mathbf{w}^{ \prime }\right ) =\rho \left (\mathbf{w}\mathbf{}^{ \prime }l_{0}\right )l_{0}^{3} =\rho \left (\mathbf{x}^{ \prime }\right )l_{0}^{3}$, where
$\rho (\mathbf{x}^{ \prime })$
is the standard mass density in
$\mbox{kg}\thinspace \mbox{m}^{ -3}$). We will use these dimensionless coordinates
$\mathbf{w}$
and
$\mathbf{w}^{ \prime }$
in the following, since they are more convenient to describe fractal media and they ensure
dimensional correctness of all physical equations.

It should also be noted that in Eq. (\ref{eq3.8}) the space dimension
$D$
might be a function of the field point coordinate
$\mathbf{w}$. On the contrary, we considered the scale length
$l_{0}$
as a constant for the particular source mass distribution being considered. In Sect. \ref{sect::galactic} we will discuss these choices in the
context of different possible galactic structures.

The ``volume'' integral over
$V_{D}$
can be performed by using techniques of multi-variable integration over a fractal metric
space
$W \subset \mathbb{R}^{3}$
\cite{bookTarasov,bookZubair,Tarasov:2014fda,TARASOV2015360}. Let's assume that
$W =W_{1} \times W_{2} \times W_{3}$, where each metric set
$W_{i}$
($i =1 ,2 ,3$) has Hausdorff measure
$\mu _{i}(W_{i})$
and dimension
$\alpha _{i}$. The Hausdorff measure for the product set
$W$
can be defined as
$\mu _{H}(W) =(\mu _{1} \times \mu _{2} \times \mu _{3})(W) =\mu _{1}(W_{1})\mu _{2}(W_{2})\mu _{3}(W_{3})$
and the overall fractal dimension is
$D =\alpha _{1} +\alpha _{2} +\alpha _{3}$. Applying Fubini's theorem we have:

\begin{gather}\int _{W}f(x_{1} ,x_{2} ,x_{3})d\mu _{H} =\int _{W_{1}}\int _{W_{2}}\int _{W_{3}}f(x_{1} ,x_{2} ,x_{3})d\mu _{1}(x_{1})d\mu _{2}(x_{2})d\mu _{3}(x_{3}) , \label{eq3.9} \\
d\mu _{i}(x_{i}) =\frac{\pi ^{\alpha _{i}/2}}{\Gamma (\alpha _{i}/2)}\left \vert x_{i}\right \vert ^{\alpha _{i} -1}dx_{i} ,\ i =1 ,2 ,3. \nonumber \end{gather}

It is straightforward to check that the integral defined in Eq. (\ref{eq3.9}) when applied to a function
$f(x_{1} ,x_{2} ,x_{3}) =f(r)$, in standard spherical coordinates
$(r ,\theta  ,\varphi )$, yields the expression in Eq. (\ref{eq3.2}). In fact,
from the
standard relations between rectangular and spherical coordinates and using the definitions
for the
the differentials in the second line of Eq. (\ref{eq3.9}), we have:
$d\mu _{1}d\mu _{2}d\mu _{3} =\frac{\pi ^{\alpha _{1}/2}}{\Gamma \left (\alpha _{1}/2\right )}\frac{\pi ^{\alpha _{2}/2}}{\Gamma \left (\alpha _{2}/2\right )}\frac{\pi ^{\alpha _{3}/2}}{\Gamma \left (\alpha _{3}/2\right )}r^{\alpha _{1} +\alpha _{2} +\alpha _{3} -1}dr\vert \sin \theta \vert ^{\alpha _{1} +\alpha _{2} -1}\left \vert \cos \theta \right \vert ^{\alpha _{3} -1}d\theta \vert \sin \varphi \vert ^{\alpha _{2} -1}\left \vert \cos \varphi \right \vert ^{\alpha _{1} -1}d\varphi $. Performing the angular integrations, simplifying the results, and using
$D =\alpha _{1} +\alpha _{2} +\alpha _{3}$
leads to the result in Eq. (\ref{eq3.2}).

If we used instead cylindrical coordinates
$(r ,\varphi  ,z)$
with a function
$f(x_{1} ,x_{2} ,x_{3}) =f(r ,z)$, we would obtain:

\begin{equation}\int _{W}fd\mu _{H} =\frac{2\pi ^{D/2}}{\Gamma [(\alpha _{r} +\alpha _{\varphi })/2]\Gamma \left (\alpha _{z}/2\right )}\int r^{\alpha _{r} +\alpha _{\varphi } -1}dr\int f(r ,z)\left \vert z\right \vert ^{\alpha _{z} -1}dz , \label{eq3.10}
\end{equation}
with
$D =\alpha _{r} +\alpha _{\varphi } +\alpha _{z}$.

From Eq. (\ref{eq3.8}), we can also re-derive the
original Eq. (\ref{eq3.5}) by using
$\widetilde{\rho }(\mathbf{w}^{ \prime }) =\widetilde{m}_{(D)}\delta ^{\left (D\right )}(\mathbf{w}^{ \prime } -0)$
where the D-dimensional Dirac delta function is defined as
$\int _{V_{D}}f(\mathbf{w}^{ \prime })\delta ^{\left (D\right )}(\mathbf{w}^{ \prime } -\mathbf{w})d^{D}\mathbf{w}^{ \prime } =f(\mathbf{w})$, for any function
$f$
over
$V_{D} \subset \mathbb{R}^{3}$. As in standard Newtonian gravity, it is possible to introduce a gravitational potential
$\phi \left (\mathbf{w}\right )$
computed as:

\begin{gather}\phi (\mathbf{w}) = -\frac{2\pi ^{1 -D/2}\Gamma (D/2)G}{\left (D -2\right )l_{0}^{2}}{\displaystyle\int _{V_{D}}}\frac{d\widetilde{m}_{(D)}}{(\left \vert \mathbf{x} -\mathbf{x}^{ \prime }\right .\vert /l_{0})^{D -2}} = -\frac{2\pi ^{1 -D/2}\Gamma (D/2)G}{\left (D -2\right )l_{0}^{2}}{\displaystyle\int _{V_{D}}}\frac{\widetilde{\rho }(\mathbf{w}^{ \prime })}{\left \vert \mathbf{w} -\mathbf{w}^{ \prime }\right .\vert ^{D -2}}d^{D}\mathbf{w}^{ \prime };\ D \neq 2 \label{eq3.11} \\
\phi \left (\mathbf{w}\right ) =\frac{2G}{l_{0}^{2}}{\displaystyle\int _{V_{2}}}d\widetilde{m}_{(2)}\ln (\left \vert \mathbf{x} -\mathbf{x}^{ \prime }\right \vert /l_{0}) =\frac{2G}{l_{0}^{2}}{\displaystyle\int _{V_{2}}}\widetilde{\rho }\left (\mathbf{w}^{ \prime }\right )\ln \left \vert \mathbf{w} -\mathbf{w}^{ \prime }\right .\vert d^{2}\mathbf{w}^{ \prime };\ D =2 \nonumber \end{gather}
with
$\phi (\mathbf{w})$
and
$\mathbf{g}(\mathbf{w})$
connected through
$\mathbf{g}(\mathbf{w}) = - \nabla _{D}\phi (\mathbf{w})$, where
$ \nabla _{D}$
represents the Del (nabla) operator in D dimensions.\protect\footnote{ Since
$ \nabla _{D}$
and all the other operators introduced later in this section will be defined in terms
of dimensionless coordinates, the physical dimensions for the gravitational potential
$\phi $
defined in Eq. (\ref{eq3.11}) are the same as
those for the gravitational field
$\mathbf{g}$, i.e., both quantities will be measured in $\mbox{m}\thinspace \mbox{s}^{ -2}$.} This general form of the scalar potential can also be derived directly by
solving the D-dimensional Laplace/Poisson equations ($ \nabla _{D}^{2}\phi (\mathbf{w}) =0$
and
$ \nabla _{D}^{2}\phi (\mathbf{w}) =\frac{4\pi G}{l_{0}^{2}}\widetilde{\rho }(\mathbf{w})$, respectively), as shown in Appendix \ref{sect::appendix_multipole} at the end of this paper.

 There is a fair amount of latitude in the definition of the
$ \nabla _{D}$-Del operator and of the other first and second order operators for the case of non-integer
dimensional spaces (see \cite{Tarasov:2014fda,TARASOV2015360} for general reviews). The axiomatic bases for spaces with non-integer dimension
were introduced by Stillinger \cite{doi:10.1063/1.523395} and Wilson
\cite{Wilson:1972cf}, then refined by Palmer and
Stavrinou \cite{Palmer_2004}. Following Eq. (\ref{eq3.2}), a radial Laplacian operator,
$ \nabla _{D}^{2}f(r) =\frac{1}{r^{D -1}}\frac{d}{dr}\left (r^{D -1}\frac{df}{dr}\right ) =\frac{d^{2}f}{dr^{2}} +\frac{(D -1)}{r}\frac{df}{dr}$, for a scalar field
$f$
can be derived \cite{doi:10.1063/1.523395}, from which the D-dimensional Laplacian in spherical coordinates follows:

\begin{gather} \nabla _{D}^{2}f =\frac{1}{r^{D -1}}\frac{ \partial }{ \partial r}\left (r^{D -1}\frac{ \partial f}{ \partial r}\right ) +\frac{1}{r^{2}\sin ^{D -2}\theta }\frac{ \partial }{ \partial \theta }\left (\sin ^{D -2}\theta \frac{ \partial f}{ \partial \theta }\right ) +\frac{1}{r^{2}\sin ^{2}\theta \sin ^{D -3}\varphi }\frac{ \partial }{ \partial \varphi }\left (\sin ^{D -3}\varphi \frac{ \partial f}{ \partial \varphi }\right ) \label{eq3.12} \\
 =\left [\frac{ \partial ^{2}f}{ \partial r^{2}} +\frac{(D -1)}{r}\frac{ \partial f}{ \partial r}\right ] +\frac{1}{r^{2}}\left [\frac{ \partial ^{2}f}{ \partial \theta ^{2}} +\frac{\left (D -2\right )}{\tan \theta }\frac{ \partial f}{ \partial \theta }\right ] +\frac{1}{r^{2}\sin ^{2}\theta }\left [\frac{ \partial ^{2}f}{ \partial \varphi ^{2}} +\frac{\left (D -3\right )}{\tan \varphi }\frac{ \partial f}{ \partial \varphi }\right ] , \nonumber \end{gather}
which obviously reduces to the standard expression for
$D =3$.

 Following Tarasov \cite{Tarasov:2014fda,TARASOV2015360}, we will extend the definition for the
divergence of a vector field
$\mathbf{F} =F_{r}\widehat{\mathbf{r}} +F_{\theta }\widehat{\boldsymbol{\theta }} +F_{\varphi }\widehat{\boldsymbol{\varphi }}$
in spherical coordinates as follows:

\begin{gather}div_{D}\mathbf{F} = \nabla _{D} \cdot \mbox{}\mathbf{F} =\left [\frac{1}{r^{D -1}}\frac{ \partial }{ \partial r}\left (r^{D -1}F_{r}\right )\right ]\widehat{\mathbf{r}} +\left [\frac{1}{r\sin ^{D -2}\theta }\frac{ \partial }{ \partial \theta }\left (\sin ^{D -2}\theta F_{\theta }\right )\right ]\widehat{\boldsymbol{\theta }} +\left [\frac{1}{r\sin \theta \sin ^{D -3}\varphi }\frac{ \partial }{ \partial \varphi }\left (\sin ^{D -3}\varphi F_{\varphi }\right )\right ]\widehat{\boldsymbol{\varphi }} \label{eq3.13} \\
 =\left [\frac{ \partial F_{r}}{ \partial r} +\frac{(D -1)}{r}F_{r}\right ]\widehat{\mathbf{r}} +\frac{1}{r}\left [\frac{ \partial F_{\theta }}{ \partial \theta } +\frac{\left (D -2\right )}{\tan \theta }F_{\theta }\right ]\widehat{\boldsymbol{\theta }} +\frac{1}{r\sin \theta }\left [\frac{ \partial F_{\varphi }}{ \partial \varphi } +\frac{\left (D -3\right )}{\tan \varphi }F_{\varphi }\right ]\widehat{\boldsymbol{\varphi }} , \nonumber \end{gather}
and assume that the definitions for gradient and curl are not affected by the fractional
dimension of the space, thus defining
$\ensuremath{\operatorname*{grad}}_{D}f = \nabla _{D}f \equiv  \nabla f$
and
$\ensuremath{\operatorname*{curl}}_{D}\mathbf{F} = \nabla _{D} \times \mathbf{u} \equiv  \nabla  \times \mathbf{u}$.\protect\footnote{ It should be noted that alternative definitions for
$ \nabla _{D}$
and for the other vector operators exist in the literature. For example \cite{bookZubair}, from the general definition of the Laplacian
$ \nabla _{D}^{2} =\left [\frac{ \partial ^{2}}{ \partial x_{1}^{2}} +\frac{\alpha _{1} -1}{x_{1}}\frac{ \partial }{ \partial x_{1}}\right ] +\left [\frac{ \partial ^{2}}{ \partial x_{2}^{2}} +\frac{\alpha _{2} -1}{x_{2}}\frac{ \partial }{ \partial x_{2}}\right ] +\left [\frac{ \partial ^{2}}{ \partial x_{3}^{2}} +\frac{\alpha _{3} -1}{x_{3}}\frac{ \partial }{ \partial x_{3}}\right ]$, with
$D =\alpha _{1} +\alpha _{2} +\alpha _{3}$
and
$0 <\alpha _{i} \leq 1$,
$i =1 ,2 ,3$, the Del operator can be defined as:$ \nabla _{D} =\sqrt{\left [\frac{ \partial ^{2}}{ \partial x_{1}^{2}} +\frac{\alpha _{1} -1}{x_{1}}\frac{ \partial }{ \partial x_{1}}\right ]}\widehat{\mathbf{x}}_{1} +\sqrt{\left [\frac{ \partial ^{2}}{ \partial x_{2}^{2}} +\frac{\alpha _{2} -1}{x_{2}}\frac{ \partial }{ \partial x_{2}}\right ]}\widehat{\mathbf{x}}_{2} +\sqrt{\left [\frac{ \partial ^{2}}{ \partial x_{3}^{2}} +\frac{\alpha _{3} -1}{x_{3}}\frac{ \partial }{ \partial x_{3}}\right ]}\widehat{\mathbf{x}}_{3} \simeq \left (\frac{ \partial \ \ \ }{ \partial x_{1}} +\frac{1}{2}\frac{\alpha _{1} -1}{x_{1}}\right )\widehat{\mathbf{x}}_{1} +\left (\frac{ \partial \ \ \ }{ \partial x_{2}} +\frac{1}{2}\frac{\alpha _{2} -1}{x_{2}}\right )\widehat{\mathbf{x}}_{2} +\left (\frac{ \partial \ \ \ }{ \partial x_{3}} +\frac{1}{2}\frac{\alpha _{3} -1}{x_{3}}\right )\widehat{\mathbf{x}}_{3}$, where the final approximation is obtained by using a binomial series expansion and
neglecting higher-order derivatives and higher powers of the coordinates in the denominators. This
approximated formula is valid only in the far-field region (i.e., for
$\left \vert x_{1}\right \vert  ,\left \vert x_{2}\right \vert  ,\left \vert x_{3}\right \vert  \gg 1$), but can be used to define all the other D-dimensional vector operators, such as
$\ensuremath{\operatorname*{grad}}_{D}f \equiv  \nabla _{D}f$,
$\ensuremath{\operatorname*{div}}_{D}\mathbf{F} \equiv  \nabla _{D} \cdot \mathbf{F}$, and
$\ensuremath{\operatorname*{curl}}_{D}\mathbf{F} \equiv  \nabla _{D} \times \mathbf{F}$, for scalar functions
$f =f(x_{1} ,x_{2} ,x_{3})$
and vector functions
$\mathbf{F} =\mathbf{F}(x_{1} ,x_{2} ,x_{3})$. With these definitions, the relation
$ \nabla _{D}^{2}f =\ensuremath{\operatorname*{div}}_{D}\ensuremath{\operatorname*{grad}}_{D}f = \nabla _{D} \cdot  \nabla _{D}f$
is only approximately true.
}

Adopting the above definitions, it is easy to check that the relation
$ \nabla _{D}^{2}f =\ensuremath{\operatorname*{div}}_{D}\ensuremath{\operatorname*{grad}}_{D}f = \nabla _{D} \cdot  \nabla _{D}f$
holds without any approximation, and that Eqs. (\ref{eq3.8}) and (\ref{eq3.11}) are connected by
$\mathbf{g}(\mathbf{w}) = - \nabla _{D}\phi (\mathbf{w})$, which actually follows from
$ \nabla _{D}\left (\frac{1}{\left \vert \mathbf{w} -\mathbf{w}\mathbf{}^{ \prime }\right \vert ^{D -2}}\right ) \equiv  \nabla \left (\frac{1}{\left \vert \mathbf{w} -\mathbf{w}\mathbf{}^{ \prime }\right \vert ^{D -2}}\right ) = -\left (D -2\right )\frac{\mathbf{w} -\mathbf{w}\mathbf{}^{ \prime }}{\left \vert \mathbf{w} -\mathbf{w}\mathbf{}^{ \prime }\right \vert ^{D}}$
(for
$D \neq 2$
)\protect\footnote{ For
$D =2$, we have
$ \nabla _{2}(\ln \left \vert \mathbf{w} -\mathbf{w}^{ \prime }\right \vert ) \equiv  \nabla (\ln \left \vert \mathbf{w} -\mathbf{w}^{ \prime }\right \vert ) =\frac{\mathbf{w} -\mathbf{w}\mathbf{}^{ \prime }}{\left \vert \mathbf{w} -\mathbf{w}\mathbf{}^{ \prime }\right \vert ^{2}}$.
} with the derivatives taken with respect to
$\mathbf{w}$, while
$\mathbf{w}^{ \prime }$
is kept constant.\protect\footnote{ The operators defined in Eqs. (\ref{eq3.12}) and (\ref{eq3.13}) as well as the gradient and
curl, are now acting on functions of dimensionless spherical coordinates
$w =\left (w_{r} ,w_{\theta } ,w_{\varphi }\right ) \equiv \left (r/l_{0} ,\theta  ,\varphi \right )$.
} Therefore, the ``fractional'' gravitational field
$\mathbf{g}(\mathbf{w})$
in Eq. (\ref{eq3.8}) can be considered to be
``conservative'' as it verifies
$ \nabla _{D} \times \mathbf{g} =0$, or
$\oint \mathbf{g} \cdot d^{\alpha }\mathbf{l} =0$, where
$d^{\alpha }\mathbf{l}$
is an appropriate infinitesimal fractional line element of dimension
$\alpha $
($0 <\alpha  \leqslant 1$). The connection between gravitational field and potential can also be expressed in
integral form as
$\Delta \Phi  = -\int \mathbf{g} \cdot d^{\alpha }\mathbf{l}$.

 If we take the fractional divergence of Eq. (\ref{eq3.8}) and use the fractional relation
$ \nabla _{D} \cdot \genfrac{(}{)}{}{}{\mathbf{w} -\mathbf{w}^{ \prime }}{\left \vert \mathbf{w} -\mathbf{w}^{ \prime }\right \vert ^{D}} =\frac{2\pi ^{D/2}}{\Gamma (D/2)}\delta ^{D}(\mathbf{w} -\mathbf{w}^{ \prime })$, we obtain the differential form of the fractional gravitational Gauss's law:

\begin{equation} \nabla _{D} \cdot \mathbf{g}\left (\mathbf{w}\right ) = -\frac{4\pi G}{l_{0}^{2}}\widetilde{\rho }\left (\mathbf{w}\right ) . \label{eq3.14}
\end{equation}
Further integration of this law over a fractional volume
$V_{D}$
of dimension
$D$
($1 <D \leq 3$) and using a fractional version of the divergence theorem,
$\oint _{S_{d}}\mathbf{g}(\mathbf{w}) \cdot d^{d}\mathbf{a} =\int _{V_{D}} \nabla _{D} \cdot \mathbf{g}(\mathbf{w})d^{D}\mathbf{w}$, involving a hypersurface
$S_{d}$
of  fractional dimension
$d =D -1$
($0 <d \leq 2$) and infinitesimal surface area
$d^{d}\mathbf{a}$, yields the integral form of the fractional Gauss theorem:

\begin{equation}\Phi _{g}^{(D)} =\oint _{S_{d}}\mathbf{g}(\mathbf{w}) \cdot d^{d}\mathbf{a} = -\frac{4\pi G}{l_{0}^{2}}\int _{V_{D}}\mathbf{}\widetilde{\rho }(\mathbf{w})d^{D}\mathbf{w} = -\frac{4\pi G}{l_{0}^{2}}\widetilde{M}_{(D)} . \label{eq3.15}
\end{equation}

 This last equation generalizes our previous Eq. (\ref{eq3.4}), where now
$\widetilde{M}_{\left (D\right )}$
represents the total mass ``enclosed'' by the hypersurface
$S_{d}$.\protect\footnote{ In Eq. (\ref{eq3.15}), we use
dimensionless coordinates
$\mathbf{w}$, so that the flux
$\Phi _{g}^{(D)}$
has the same dimensions of the field
$\mathbf{g}$ ($\mbox{m}\thinspace \mbox{s}^{ -2}$). In Eq. (\ref{eq3.4}) we were still using
standard coordinates and the flux
$\Phi _{g}^{(D)}$
was measured in $\mbox{m}^{D}\thinspace \mbox{s}^{ -2}$. The two equations are equivalent also in view of the redefinition of the fractional mass
as
$m_{(D)} =\widetilde{m}_{(D)}/l_{0}^{3 -D}$.
} In cases of high symmetry (spherical, cylindrical, etc.) it can be used to
determine the gravitational field
$\mathbf{g}$
as it is usually done in standard Newtonian gravity. Combining together the expression for
the divergence of the field with the relation between field and potential given above we can also
obtain a fractional Poisson equation:

\begin{equation} \nabla _{D}^{2}\phi \left (\mathbf{w}\right ) =\frac{4\pi G}{l_{0}^{2}}\widetilde{\rho }\left (\mathbf{w}\right ) . \label{eq3.16}
\end{equation}

In Appendix \ref{sect::appendix_multipole}, we will solve the related D-dimensional Laplace equation
$ \nabla _{D}^{2}\phi \left (\mathbf{w}\right ) =0$, for regions where
$\widetilde{\rho }\left (\mathbf{w}\right ) =0$, and derive the corresponding multipole expansion. Other formulas of Newtonian
gravity and potential theory can also be generalized to fractional dimensional cases, but this
analysis would go beyond the scope of this paper and will be left to further studies. In the next
section, we will apply the ideas developed above to some fundamental models of galactic structures.

We emphasize again that all the NFDG\  formulas presented in this section imply the use of linear operators acting on a metric space of fractional dimension $D$, as opposed to using fractional vector calculus \cite{2006PhyA..367..181M,2020arXiv200507686D} where vector operators such as gradient, divergence, curl, etc., are defined in terms of fractional integro-differential operators. Our choice was based on the assumption that galactic structures might behave like fractal media of non-integer dimension $D$; therefore, a \textit{Non-integer-dimensional space approach}, as described by Tarasov (\cite{Tarasov:2014fda} and references therein), is more appropriate for this case. As already mentioned in this section, these methods based on integration and differentiation for non integer-dimensional spaces are well developed  \cite{doi:10.1063/1.523395,Wilson:1972cf,Palmer_2004}, and widely used also in quantum field theory \cite{Bollini:1972ui,tHooft:1972tcz,Wilson:1972cf}.

We also remark that our hypothesis that galaxies might behave like fractal media with $D \neq 3$, does not seem to be connected to extra-dimension theories and related searches \cite{2018PhRvD..98c0001T} (large extra dimensions, warped extra dimensions, etc.), or to the limits on the number of space-time dimensions imposed by the propagation of gravitational waves \cite{2018PhRvD..97f4039V,2018JCAP...07..048P,2019PhRvD.100h4050K}, since our model proposes an effective modification of the law of gravity only for galactic structures which possibly behave as fractal media, and not for single (or binary) astrophysical objects (stars, black holes, etc.) and related gravitational wave emissions. 

As for the physical origin of the NFDG variable dimension $D$ (with $2 <D \leq 3$, for the spherical structures analyzed in this paper), we are unable to propose any explanation until our methods are applied to individual galaxies and related data. This will be done in an upcoming work \cite{Varieschi:2020dnd}, which will apply NFDG to axially-symmetric structures and will include detailed fitting of individual disk galaxies.

\section{\label{sect::galactic}
Galactic models
}
From the analysis presented in Sect. \ref{sect::gauss}, it is clear that our NFDG is intended as a modification of the law of gravity and
not of the law of inertia and Newtonian dynamics. However, it is possible that similar modifications
might also affect the other forces in nature (e.g., the electromagnetic force, since the theory
outlined in Sect. \ref{sect::gauss} was originally
introduced for the electromagnetic field \cite{Boito:2018rdh,McDonald:2018,bookZubair}).\protect\footnote{ Fractional electrodynamics has already been applied successfully to some
fractal media and materials (see again \cite{bookTarasov,bookZubair} and references therein), thus showing that
classical forces can be affected by the dimensionality of space.
}

 In this way, our interpretation is somewhat in between the two original forms of
the MOND theory outlined at the beginning of Sect. \ref{sect:MOND}. Nevertheless, we will assume that Newtonian dynamics (i.e., Newton's laws of motion
and related consequences, such as conservation principles, etc.) is not affected in any way by our
fractional generalizations. A test object, subject to a fractional gravitational field, such as the
one described by Eq. (\ref{eq3.8}), will still move in a
(classical)
$3 +1$
space-time, thus obeying standard laws of dynamics.

 The structure and the dynamics of galaxies are described in detail in the seminal
monograph by Binney and Tremaine \cite{2008gady.book.....B}. In the
following sub-sections, we will apply our NFDG to some fundamental galactic structures and connect
our
results with the empirical MOND predictions outlined in Sect. \ref{sect:MOND}. In this paper, we will limit ourselves to cases of spherical
symmetry, leaving other types of symmetries and geometries to future work \cite{Varieschi:2020dnd}.

\subsection{\label{subsect:spherical}
Spherical symmetry
}
 The gravitational field
$\mathbf{g}(w)$, due to a spherically symmetric source mass distribution
$\widetilde{\rho }(w^{ \prime })$, in a fractal space of dimension
$D(w^{})$ depending on the distance from the center of the coordinate system, can be computed
directly by
generalizing the standard Newtonian derivation based on the computation of the field
due to an infinitesimal spherical shell.

 The full derivation is presented in Appendix \ref{sect:appendix_spherical}. There we show how, starting from Eq. (\ref{eq3.8}), we obtain the
following general expression (same as Eq. (\ref{eq7.3}) in
Appendix \ref{sect:appendix_spherical}):

\begin{equation}\mathbf{g}_{obs}(w) = -\frac{4\pi G}{l_{0}^{2}w^{D\left (w\right ) -1}}{\displaystyle\int _{0}^{w}}\tilde{\rho }\left (w^{ \prime }\right )w^{ \prime ^{D\left (w\right ) -1}}dw^{ \prime }\overset{}{\widehat{\mathbf{w}} ,} \label{eq4.1}
\end{equation}
which was proved for
$1 \leq D \leq 3$.

 In the previous equation, we also denoted the gravitational field as the
``observed'' one,
$\mathbf{g}_{obs}$, as opposed to the ``baryonic''
$\mathbf{g}_{bar}$:

\begin{equation}\mathbf{g}_{bar}(w) = -\frac{4\pi G}{l_{0}^{2}w^{2}}{\displaystyle\int _{0}^{w}}\tilde{\rho }\left (w^{ \prime }\right )w^{ \prime ^{2}}dw^{ \prime }\overset{}{\widehat{\mathbf{w}} ,} \label{eq4.2}
\end{equation}
for fixed dimension
$D =3$. In other words, we identify the observed and baryonic accelerations
$g_{obs}$
and
$g_{bar}$\cite{McGaugh:2016leg} with those obtained in NFDG for
variable dimension
$D$
and for fixed dimension
$D =3$, respectively.

 With these NFDG assumptions, for spherically symmetric structures, the ratio
$(g_{obs}/g_{bar})_{NFDG}$
is simply obtained from Eqs. (\ref{eq4.1}) and (\ref{eq4.2}):

\begin{equation}\genfrac{(}{)}{}{}{g_{obs}}{g_{bar}}_{NFDG}(w) =w^{3 -D\left (w\right )}\frac{{\displaystyle\int _{0}^{w}}\widetilde{\rho }\left (w^{ \prime }\right )w^{ \prime ^{D\left (w\right ) -1}}dw^{ \prime }}{{\displaystyle\int _{0}^{w}}\widetilde{\rho }\left (w^{ \prime })\right .w^{\prime ^{2}} dw^{ \prime }} , \label{eq4.3}
\end{equation}
where the dimension function
$D(w)$
needs to be determined either from experimental data or from theoretical considerations.
In this section, we will consider the first option, while possible theoretical determination of the
dimension function will be left for future work on the subject \cite{Varieschi:2020dnd}.

In general, to obtain
$D(w)$
from experimental data, without fitting any particular set of galactic data, we simply
compare the expression in Eq. (\ref{eq4.3}) with the MOND
equivalent expression from Eq. (\ref{eq2.6}),
$\genfrac{(}{)}{}{}{g_{obs}}{g_{bar}}_{MOND}\left (w\right ) =\frac{1}{1 -e^{ -\sqrt{g_{bar}\left (w\right )/g_{\dag }}}}$, or with similar expressions obtained by using the other interpolation functions in Eq.
(\ref{eq2.5}). If these two expressions are compatible, we expect to obtain
$D(w^{})$
as a continuous function with values
$D \approx 3$
in regions where Newtonian gravity holds. The dimension should then decrease
toward $D \approx 2$ in regions where the deep-MOND regime applies, following our general discussion in Sect. \ref{sect:MOND}.

\subsection{\label{subsect:pointmass}
Point-mass objects
}
 Following Eq. (\ref{eq3.2}), a point-mass
$m$
placed at the origin of a
$D$-dimensional space can be represented, in spherical coordinates, by the following mass
density:

\begin{equation}\widetilde{\rho }\left (w^{ \prime }\right ) =\frac{m\Gamma \left (D/2\right )\delta \left (w^{ \prime }\right )}{2\pi ^{D/2}w^{ \prime D -1}} , \label{eq4.4}
\end{equation}
which reduces to the standard expression
$\widetilde{\rho }\left (w^{ \prime }\right ) =\frac{m\delta \left (w^{ \prime }\right )}{4\pi w^{ \prime 2}}$
for
$D =3$. With this choice for the mass density, assuming
$a_{0} =Gm/l_{0}^{2}$
from Eq. (\ref{eq3.6}), and using the properties of
the Dirac delta functions inside the integrals in Eqs. (\ref{eq4.1})-(\ref{eq4.3}), all the relevant quantities are
easily computed.

 The
$y$
parameter from Eq. (\ref{eq2.4}) simply becomes
$y =\frac{g_{bar}}{a_{0}} =\frac{1}{w^{2}}$, from which the functions in Eq. (\ref{eq2.5})
corresponding to
$\genfrac{(}{)}{}{}{g_{obs}}{g_{bar}}_{MOND}$
are easily obtained. In particular:
\begin{gather}\nu _{n}\left (y\right ) =\left (\frac{1}{2} +\frac{1}{2}\sqrt{1 +4w^{2n}}\right )^{1/n} \label{eq4.5} \\
\widehat{\nu }_{n}\left (y\right ) =\left [1 -\exp \left ( -w^{ -n}\right )\right ]^{ -1/n} \nonumber \end{gather}
and

\begin{gather}g_{bar}\left (w\right ) =\frac{a_{0}}{w^{2}} \label{eq4.6} \\
g_{obs}\left (w\right ) =\frac{a_{0}2\pi ^{1 -D\left (w\right )/2}\Gamma \left (D\left (w\right )/2\right )}{w^{D\left (w\right ) -1}} \nonumber  \\
\genfrac{(}{)}{}{}{g_{obs}}{g_{bar}}_{NFDG}\left (w\right ) =2\pi ^{1 -D\left (w\right )/2}\Gamma \left (D\left (w\right )/2\right )w^{3 -D\left (w\right )} . \nonumber \end{gather}

In view of Eq. (\ref{eq2.6}) in Sect. \ref{sect:MOND}, the variable dimension
$D\left (w\right )$
appearing in the second and third lines of the previous equation is obtained by equating
the function
$\genfrac{(}{)}{}{}{g_{obs}}{g_{bar}}_{NFDG}\left (w\right )$
in the same Eq. (\ref{eq4.6}) with either
$\nu _{n}\left (w\right )$
or
$\widehat{\nu }_{n}\left (w\right )$ from Eq. (\ref{eq4.5}), and by solving numerically the
resulting equation.

\begin{figure}\centering 
\setlength\fboxrule{0in}\setlength\fboxsep{0.1in}\fcolorbox[HTML]{FFFFFF}{FFFFFF}{\includegraphics[ width=7.183333333333333in, height=6.724999999999999in,]{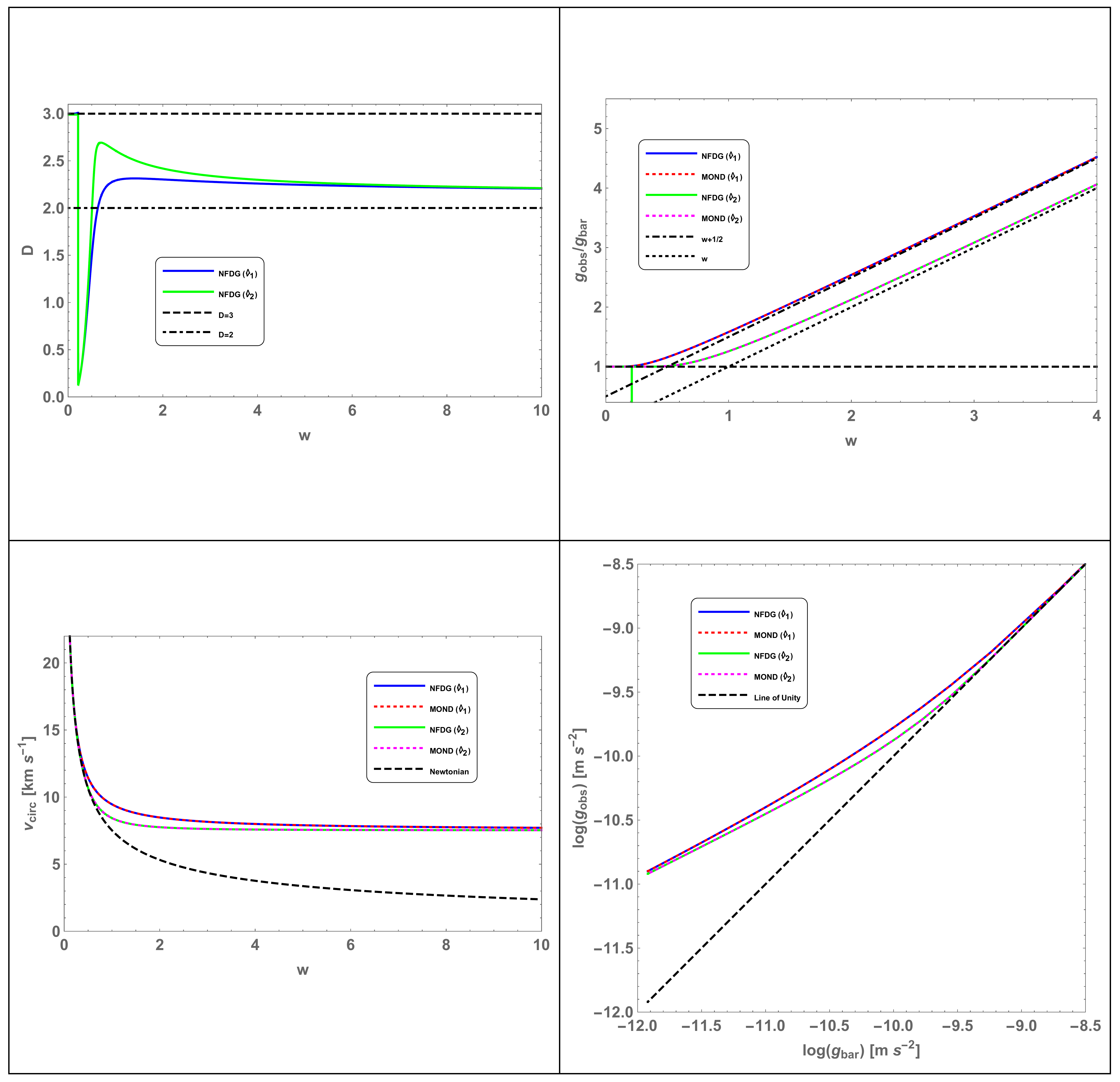}
}
\caption{Point-mass object results.
Top-left panel: NFDG variable dimension $D\left (w\right )$ for MOND interpolation functions $\widehat{\nu }_{1}$ and $\widehat{\nu }_{2}$. Other panels: comparison of NFDG\ results (solid lines) with equivalent MOND predictions (dotted lines) for the two different interpolation functions. Also shown: Newtonian behavior-Line of Unity (black-dashed lines).
}\label{figure: Point}\end{figure}

 Figure \ref{figure: Point} shows all
the results for this particular case. The top-left panel illustrates the variable dimension
$D\left (w\right )$
obtained using some of the functions in Eq. (\ref{eq4.5}).
We can see that the MOND\ functions are not fully compatible
with the NFDG analysis at low values for
$w$, where the dimension suddenly changes from
$D \approx 3$
in the initial Newtonian regime, to a low value close to zero. This is due to the fact
that the equation
$\genfrac{(}{)}{}{}{g_{obs}}{g_{bar}}_{NFDG}\left (w\right ) =\widehat{\nu }_{1}\left (w\right )$
(or
$\widehat{\nu }_{2}\left (w\right )$) admits two possible numerical solutions in this region, where one of the two becomes
progressively greater than
$D =3$ and, therefore, unphysical. In the figure we only show values
$0 <D \lesssim 3$. At higher values for the variable
$w$, the dimension
$D$
progressively decreases toward the
$D =2$
value, as it is to be expected in the deep-MOND regime, following the discussion in Sect.
\ref{sect::gauss}.

 The top-right panel in the figure shows the ratio
$\frac{g_{obs}}{g_{bar}}$
computed in two different ways:
$\genfrac{(}{)}{}{}{g_{obs}}{g_{bar}}_{NFDG}\left (w\right )$
from Eq. (\ref{eq4.6}) with the dimension
$D\left (w\right )$
obtained before, and
$\genfrac{(}{)}{}{}{g_{obs}}{g_{bar}}_{MOND}\left (w\right ) =\widehat{\nu }_{1}\left (w\right )$
(or
$\widehat{\nu }_{2}\left (w\right )$), simply using the two MOND functions in Eq. (\ref{eq4.5}). In both cases, the NFDG plots (solid lines) match the MOND ones (dotted lines), showing a ratio
$\frac{g_{obs}}{g_{bar}} \simeq 1$
at low
$w$
within the Newtonian regime. In the deep-MOND high-$w$
range we have instead
$\frac{g_{obs}}{g_{bar}} \sim w +\frac{1}{2}$, or $\frac{g_{obs}}{g_{bar}} \sim w$, for the two cases related to $\widehat{\nu }_{1}$ and $\widehat{\nu }_{2}$ respectively, as expected in the MOND model. The results shown in these two top panels of figure \ref{figure: Point} are independent of the mass
$m$
of the point-like object, and were obtained by using only the
$n =1 ,2$
values for the general MOND function $\widehat{\nu }_{n}$ in Eq. (\ref{eq2.5}). Using
$n >2$
values for the same function $\widehat{\nu }_{n}$ does not yield results which are much different from the
$n =2$
ones.\protect\footnote{ Using MOND functions
$\nu _{n}$,
instead of
$\widehat{\nu }_{n}$,
yields very similar results for all values of $n$. Therefore, we have considered only the $\widehat{\nu }_{n}$ family of MOND interpolation functions in this work.}

 The bottom-left panel shows circular velocity plots corresponding to the
previously analyzed cases, and compared with the purely Newtonian case. For this panel, as well as for the
bottom-right one, we have assumed a mass
$m \approx 2 \times 10^{5}M_{ \odot } \approx 4 \times 10^{35}\thinspace \mbox{kg}$
, typical of globular clusters in our Galaxy \cite{2008gady.book.....B}, but the results would be similar for any other choice of mass $m$. In this panel the NFDG circular speeds are
computed as
$v_{circ} =\sqrt{g_{obs}\left (w\right )w\thinspace l_{0}}/10^{3}$$\left [\mbox{km}\thinspace \mbox{s}^{ -1}\right ]$
with
$l_{0} \approx \sqrt{\frac{Gm}{a_{0}}} \simeq 4.72 \times 10^{17}\mbox{m}$. The MOND circular speeds are computed as
$v_{circ} =\sqrt{g_{bar}\left (w\right )\widehat{\nu }_{1}\left (w\right )w\thinspace l_{0}}/10^{3}\left [\mbox{km}\thinspace \mbox{s}^{ -1}\right ]$
(or using
$\widehat{\nu }_{2}$
instead of
$\widehat{\nu }_{1}$) and the purely Newtonian speed is
$v_{circ} =\sqrt{g_{bar}\left (w\right )w\thinspace l_{0}}/10^{3}\left [\mbox{km}\thinspace \mbox{s}^{ -1}\right ]$. As seen from the panel, there is perfect agreement between the respective ($\widehat{\nu }_{1}$
or
$\widehat{\nu }_{2}$) NFDG and MOND cases, showing the expected flattening of the circular speed plots at
high-$w$, as opposed to the standard Newtonian decrease of circular speed with radial distance ($v_{circ} \sim 1/\sqrt{w}$ for Newtonian behavior).

 Finally, the bottom right panel is similar to the
$\log \left (g_{obs}\right )$
vs.
$\log \left (g_{bar}\right )$
plots widely used in the literature (see Fig. 3 in Ref. \cite{McGaugh:2016leg} or the figures in Ref. \cite{Lelli:2017vgz}) to illustrate the validity of the general MOND-RAR relation in Eq.
(\ref{eq2.6}). Compared to the \textit{Line of Unity},
representing the purely Newtonian case, there is again perfect agreement over several orders of
magnitude between plots obtained with our NFDG\ model, using
$g_{obs}\left (w\right )$
and
$g_{bar}\left (w\right )$
from Eq. (\ref{eq4.6}), and MOND plots where
$g_{obs}\left (w\right ) =\widehat{\nu }_{1}\left (w\right )g_{bar}\left (w\right )$
(or
$g_{obs}\left (w\right ) =\widehat{\nu }_{2}\left (w\right )g_{bar}\left (w\right )$).

 Apart from the discontinuity at low-$w$
in the top-left panel, which will disappear in the next two cases analyzed in this
section, the study of the point-mass case already shows that the variable-dimension effect of NFDG
can be equivalent to the MOND-RAR model. In the next two sub-sections we will confirm this result
using two other spherically symmetric cases.

\subsection{\label{subsect:homogeneous}
Homogeneous sphere
}
 A homogeneous sphere of radius
$R$
and total mass
$M$
will have constant mass density
$\rho  =\frac{3M}{4\pi R^{3}}$
inside the sphere and
$\varrho  =0$
outside. Rescaling all distances as usual,
$w^{ \prime } =r/l_{0}$,
$W =R/l_{0}$, and using a Heaviside step function
$H$
we can write the rescaled mass density as:

\begin{equation}\widetilde{\rho }(w^{ \prime }) =\frac{3M}{4\pi W^{3}}H(W -w^{ \prime }) . \label{eq4.7}
\end{equation}

 We then follow the same procedure outlined in the previous subsection \ref{subsect:pointmass}. The
$y$
parameter is easily computed as:

\begin{equation}y =\frac{g_{bar}}{a_{0}} =\frac{1}{W^{3}w^{2}}\left [W^{3} +\left (w^{3} -W^{3}\right )H\left (W -w\right )\right ] , \label{eq4.8}
\end{equation}
from which the MOND functions
$\widehat{\nu }_{1}$
and
$\widehat{\nu }_{2}$
also follow. Functions
$g_{obs}\left (w\right )$,
$g_{bar}\left (w\right )$, and their ratio
$\genfrac{(}{)}{}{}{g_{obs}}{g_{bar}}_{NFDG}\left (w\right )$
will follow from Eqs. (\ref{eq4.1})-(\ref{eq4.3}), with the dimension
$D\left (w\right )$
obtained by solving numerically the equation
$\genfrac{(}{)}{}{}{g_{obs}}{g_{bar}}_{NFDG}\left (w\right ) =\widehat{\nu }_{1}\left (w\right )$
(or
$\widehat{\nu }_{2}\left (w\right )$). The total mass is chosen to be the same as the one used in the point-mass case, i.e.,
$M \approx 2 \times 10^{5}M_{ \odot } \approx 4 \times 10^{35}\mbox{kg}$, typical of globular clusters. The other physical parameter
$W =R/l_{0}$
is chosen as
$W =0.1$. We will use the same value in the next sub-section for the Plummer model
(see Sect. \ref{subsect:plummer}) and this choice will be justified in terms of the astrophysical parameters of globular clusters.

\begin{figure}\centering 
\setlength\fboxrule{0in}\setlength\fboxsep{0.1in}\fcolorbox[HTML]{FFFFFF}{FFFFFF}{\includegraphics[ width=7.183333333333333in, height=6.724999999999999in,]{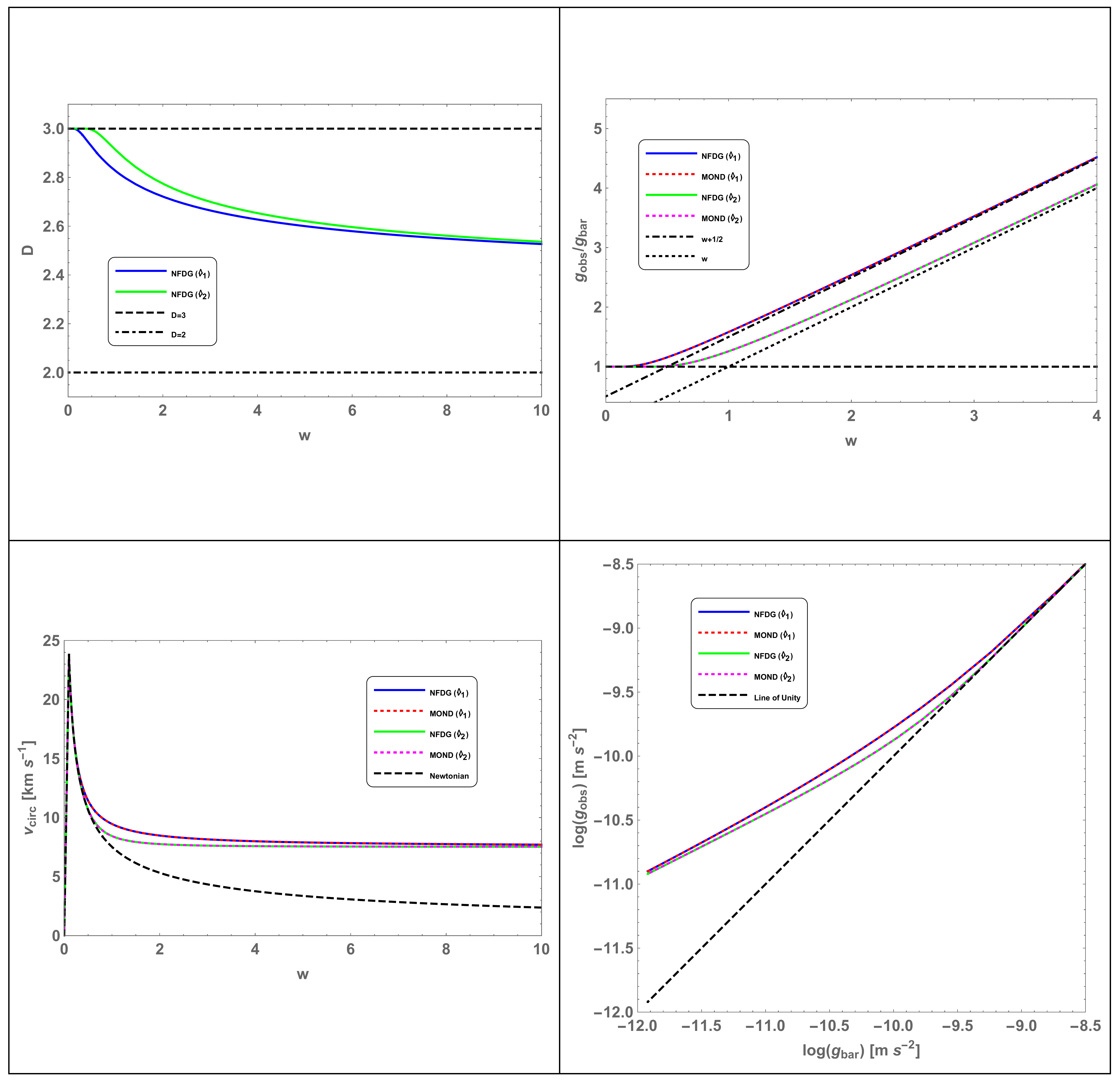}
}
\caption{Homogeneous sphere results.
Top-left panel: NFDG variable dimension $D\left (w\right )$ for MOND interpolation functions $\widehat{\nu }_{1}$ and $\widehat{\nu }_{2}$. Other panels: comparison of NFDG\ results (solid lines) with equivalent MOND predictions (dotted lines) for the two different interpolation functions. Also shown: Newtonian behavior-Line of Unity (black-dashed lines).
}\label{figure: Homogeneous}\end{figure}

 Figure \ref{figure: Homogeneous}
shows the results for this case, in the same way of figure \ref{figure: Point} previously. The top left panel illustrates the dimension functions
$D\left (w\right )$
for the two cases being considered. This time, the dimension functions are uniquely
defined and continuous over the whole range: at low-$w$
values,
$D \approx 3$
in the Newtonian regime, then the dimension continuously decreases approaching the value
$D \approx 2$
in the deep-MOND regime, as expected.

 The top-right panel shows the same two regimes, Newtonian and deep-MOND, in terms
of the
$\frac{g_{obs}}{g_{bar}}$
ratio: close to unity at low-$w$
(Newtonian) and approaching asymptotically
$w +1/2$
(or $w$) at high-$w$
(deep-MOND). Finally, the two bottom panels show the equivalent circular speed and
log-log plots, with perfect correspondence between the NFDG and MOND computations (obtained with the
same procedure outlined above for figure \ref{figure: Point}).

\subsection{\label{subsect:plummer}
Plummer model
}
 The last spherical model considered in this section is based on the Plummer
gravitational potential
$\phi \left (r\right ) = -\frac{GM}{\sqrt{r^{2} +b^{2}}}$
and related mass density
$\rho (r) =\frac{3M}{4\pi b^{3}}\left (1 +\frac{r^{2}}{b^{2}}\right )^{ -5/2}$, where
$M$
is the total mass of the system and
$b$
is the Plummer scale length \cite{2008gady.book.....B}. As
we did for the homogeneous model in the previous sub-section, we rescale all distances,
$w^{ \prime } =r/l_{0}$,
$W =b/l_{0}$, and obtain the effective mass density:

\begin{equation}\widetilde{\rho }\left (w^{ \prime }\right ) =\frac{3M}{4\pi W^{3}}\left (1 +\frac{w^{ \prime 2}}{W^{2}}\right )^{ -5/2} . \label{eq4.9}
\end{equation}

 Always assuming
$a_{0} =\frac{GM}{l_{0}^{2}}$ and using Eq. (\ref{eq4.2}), we compute
$y =\frac{g_{bar}}{a_{0}} =\frac{w}{W^{3}}\left (1 +\frac{w^{2}}{W^{2}}\right )^{ -3/2}$
and then all other quantities follow from the general equations (\ref{eq4.1})-(\ref{eq4.3}) as in the previous
cases, with the integrations performed either analytically or numerically. The Plummer model is
usually associated with spherical structures such as globular clusters, therefore we use again the
reference mass $M \approx 2 \times 10^{5}M_{ \odot } \approx 4 \times 10^{35}\mbox{kg}$, which yields
$l_{0} =\sqrt{\frac{GM}{a_{0}}} \simeq 4.72 \times 10^{17}\mbox{m}$. Since the typical half-mass radius of globular clusters in our Galaxy is
$r_{h} \approx 3\thinspace pc \approx 10^{17}\mbox{m}$ \cite{2008gady.book.....B}, and for the Plummer mass density
above
$r_{h}/b \simeq 1.30$,
we estimate
$b \simeq 7.10 \times \mbox{}10^{16}\thinspace \mbox{m}$
and thus
$W =b/l_{0} \approx 0.1$, as also used in sub-section \ref{subsect:homogeneous}.

\begin{figure}\centering 
\setlength\fboxrule{0in}\setlength\fboxsep{0.1in}\fcolorbox[HTML]{FFFFFF}{FFFFFF}{\includegraphics[ width=7.183333333333333in, height=6.724999999999999in,]{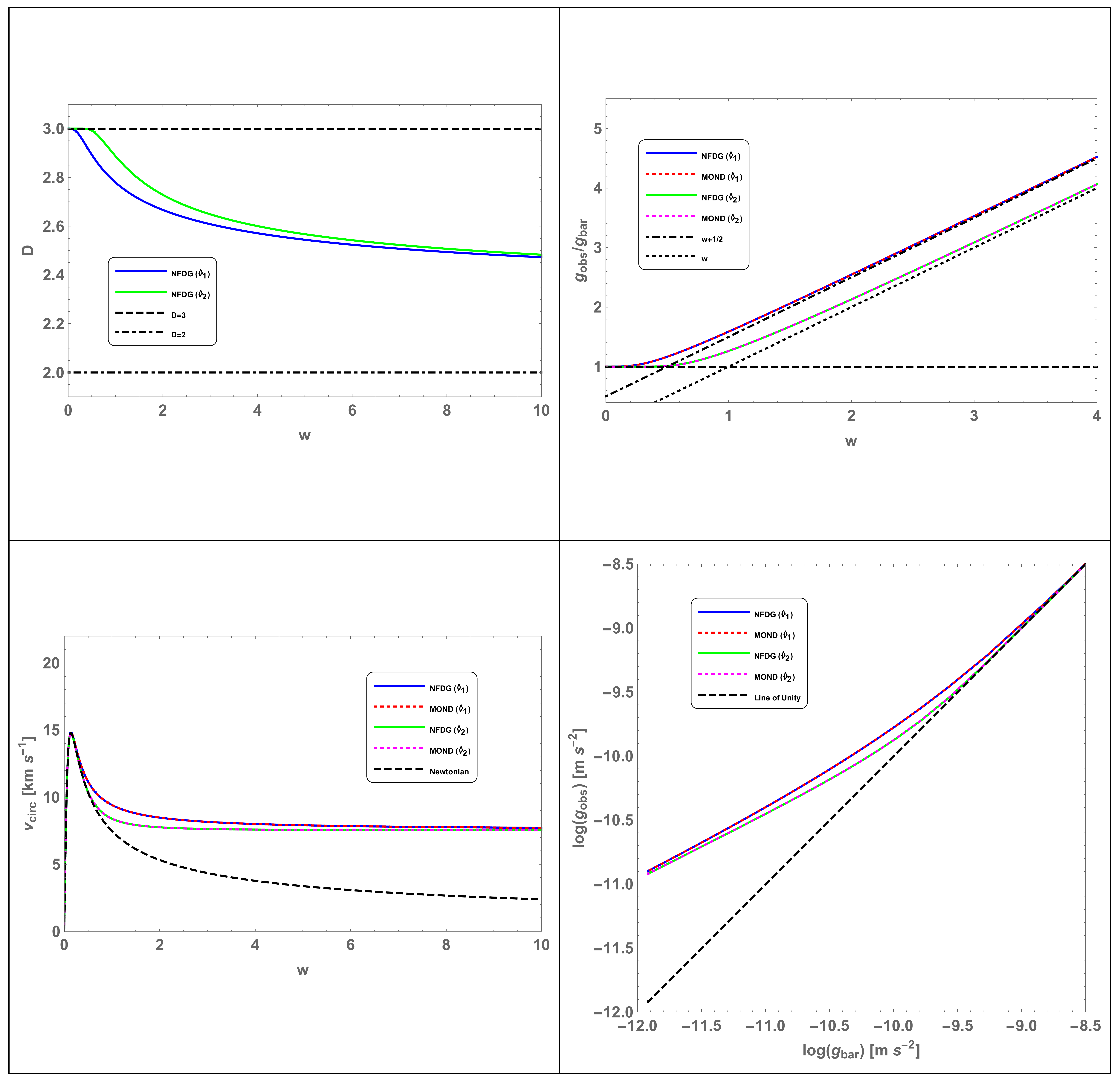}
}
\caption{Plummer model results.
Top-left panel: NFDG variable dimension $D\left (w\right )$ for MOND interpolation functions $\widehat{\nu }_{1}$ and $\widehat{\nu }_{2}$. Other panels: comparison of NFDG\ results (solid lines) with equivalent MOND predictions (dotted lines) for the two different interpolation functions. Also shown: Newtonian behavior-Line of Unity (black-dashed lines).
}\label{figure: Plummer}\end{figure}

 Figure \ref{figure: Plummer},
modeled after the first two figures, summarizes all results for this case. Once again, the top-left
panel shows dimension functions
$D\left (w\right )$
consistent with our NFDG model based on fractional gravity: the effective dimension of the
physical space decreases from the standard Newtonian
$D \approx 3$
value toward the deep-MOND
$D \approx 2$
value asymptotically.\protect\footnote{ In all these figures, we limited the range of
$w$
between
$0$
and
$10$. Plotting the top-left panels for
$w \gg 10$ would show that, in fact,
$D \rightarrow 2$
for large values of
$w$.
} This is also reflected in the top-right panel, by using the
$\frac{g_{obs}}{g_{bar}}$
ratios instead. The two bottom panels in figure \ref{figure: Plummer}, also confirm that our NFDG model can yield the same results of the standard
MOND theory, but with circular speed plots and log-log plots now explained by our variable-dimension effect, as
opposed to just an empirical MOND-RAR relation.

Although in this section we used globular clusters and related astrophysical data in our models, the NFDG analysis can be easily extended to any other spherical stellar structure, such as dwarf spheroidal galaxies or others. Additional simulations corresponding to other choices of spherical stellar structures, with different mass densities and different astrophysical data, produced results similar to those in Figs. \ref{figure: Point}-\ref{figure: Plummer} and, therefore, were not included in the current work.

\section{\label{sect::conclusion}
Conclusion
}
In this work, we outlined a possible explanation of the MOND theory and related RAR
in terms of a novel fractional-dimension gravity model. We considered the possibility that Newtonian gravity
might act on a metric space of variable dimension
$D \leq 3$, when applied to galactic scales, and developed the mathematical basis of a classical NFDG.

In particular, the MOND acceleration scale
$a_{0}$, or the equivalent RAR acceleration parameter
$g_{\dag } =1.20 \times 10^{ -10}\mbox{m}\thinspace \mbox{s}^{ -2}$, can be related to a length scale
$l_{0} \approx \sqrt{\frac{GM}{a_{0}}}$
which is naturally required for dimensional reasons when dealing with Newtonian gravity in $2<D<3$.

We have shown that, at least for some fundamental spherically-symmetric cases, our
NFDG can reproduce the same results of the MOND-RAR models, and that the deep-MOND regime can be
achieved by continuously decreasing the space dimension
$D$
toward a limiting value of
$D \approx 2$.

Future work on the subject \cite{Varieschi:2020dnd} will
be needed to test the NFDG hypothesis. This model will need to be extended to galactic structures with
axial symmetry and detailed fitting to galactic rotation curves will also need to be performed, before NFDG
can be considered a viable alternative model. Lastly, the origin of the supposed continuous
variation of the space dimension
$D$
has to be determined, possibly arising from a relativistic version of fractional-dimension gravity.

\begin{acknowledgments}This work was supported by a Faculty Fellowship awarded by the Frank R. Seaver
College of Science and Engineering, Loyola Marymount University. The author wishes to
acknowledge Dr. Ben Fitzpatrick for his advice regarding numerical computations with
Mathematica, and also thank the anonymous reviewers for their useful comments and suggestions.

\end{acknowledgments}\appendix

\section{\label{sect::appendix_multipole}
Laplace equation in D-dimensions and the multipole expansion
}
Several studies of fractional
gravity exist in the literature \cite{Muslih2007,Rousan2002,Calcagni:2009kc,Calcagni:2011kn,Calcagni:2011sz,Calcagni:2013yqa,Calcagni:2016azd,Calcagni:2018dhp} but, to our knowledge, they do not discuss a NFDG as outlined in this paper, with the possible exception of a recent paper by Giusti  \cite{2020PhRvD.101l4029G}, which appeared in the literature while our work was being finalized.

In this paper, Giusti introduces a fractional version of Newton's theory based on a fractional Poisson equation, $\left ( -\Delta \right )^{s}\Phi \left (\mathbf{x}\right ) = -4\pi G_{N}\ell ^{2 -2s}\rho \left (\mathbf{x}\right )$, where $\Delta $ is the standard Laplacian, $\Phi \left (\mathbf{x}\right )$ the modified gravitational potential, $\rho \left (\mathbf{x}\right )$ the matter density distribution, and $G_{N}$ is the standard Newtonian gravitational constant that we simply denoted by $G$ in this paper. The parameter $s$ can vary in the range $1$$ \leq s \lesssim 3/2$, where $s =1$ corresponds to Newtonian gravity, while $s =3/2$ yields MOND-like behavior. The scale length $\ell  =\frac{2}{\pi }\sqrt{\frac{G_{N}M}{a_{0}}}$ plays a similar role of our NFDG length $l_{0} \approx \sqrt{\frac{GM}{a_{0}}}$  and is used by Giusti to obtain the main empirical results of MOND, as we did by using $l_{0}$ in our Sect. \ref{sect::gauss}. In that section, we remarked that the fundamental MOND results can be recovered in NFDG by choosing any scale length proportional to the quantity $\sqrt{\frac{GM}{a_{0}}}$ (see our footnote 7), so the factor of $\frac{2}{\pi }$ appearing in Giusti's definition above is not in conflict with our NFDG results.

We also note that the fractional Poisson equation proposed by Giusti does not correspond to our Eq. (\ref{eq3.16}) in Sect. \ref{sect::gauss}, and that the potential obtained by Giusti  \cite{2020PhRvD.101l4029G} in his equation (20), for a point-like mass $M$, differs from our expression in Eq.  (\ref{eq3.11}). Giusti summarizes his main results as:

\begin{gather}\Phi _{s}\left (r\right ) = -\frac{\Gamma \left (\frac{3}{2} -s\right )}{4^{s -1}\sqrt{\pi }\Gamma \left (s\right )}\genfrac{(}{)}{}{}{\ell }{r}^{2 -2s}\frac{G_{N}M}{r}\text{, for }0 \leq s <3/2 , \label{eq6.0} \\
\Phi _{s}\left (r\right ) =\frac{2}{\pi }\frac{G_{N}M}{\ell }\log \left (r/\ell \right )\text{, for }s =3/2. \nonumber \end{gather}

In fact, assuming that the scale lengths of the two models are simply connected by $\ell  =\frac{2}{\pi }l_{0}$ and that the relation between $s$ and our fractional dimension $D$ is linear: $s =\left (5 -D\right )/2$, linking the Newtonian behavior ($s =1$, $D =3$) to the deep-MOND regime ($s =3/2$, $D =2$), it is easy to check that the NFDG and Giusti logarithmic potentials in the second lines of Eqs. (\ref{eq3.11}) and (\ref{eq6.0}) respectively, are very similar, but not identical. In our notation, Giusti's logarithmic potential becomes: $\phi \left (w\right ) =\frac{GM}{l_{0}^{2}}\ln \genfrac{(}{)}{}{}{\pi w}{2}$, compared with the NFDG logarithmic potential following from Eq. (\ref{eq3.11}): $\phi \left (w\right ) =\frac{2GM}{l_{0}^{2}}\ln \left (w\right )$, thus differing by an overall factor of $2$ (and also by an additive constant, which will not affect the resulting gravitational field).

In addition, if we rewrite in the notation used in this paper also the first line of Giusti's potential above, using all the previous assumptions, we obtain $\phi \left (w\right ) = -\genfrac{[}{]}{}{}{4^{D -3}\pi ^{5/2 -D}\Gamma \left (\frac{D}{2} -1\right )}{\Gamma \genfrac{(}{)}{}{}{5 -D}{2}}\frac{GM}{l_{0}^{2}w^{D -2}}$, which is not fully equivalent to the first line in our Eq. (\ref{eq3.11}), which can be rewritten for a point-like mass $M$ as: $\phi \left (w\right ) = -\genfrac{[}{]}{}{}{2\pi ^{1 -D/2}\Gamma \left (\frac{D}{2}\right )}{\left (D -2\right )}\frac{GM}{l_{0}^{2}w^{D -2}}$. The two functions of $D$ in the square brackets of these expressions appear to be different, but they show a remarkable similarity when plotted over the range $2 <D <3$, thus yielding almost equivalent results over this range. In other possible ranges, such as for $0 <D <2$, the two functions are not similar anymore (and Eq. \ref{eq6.0} is considered by Giusti to hold only for $2 <D <5$).

Summarizing this discussion: the overall form of the fractional potential, $\phi  \sim  -\frac{GM}{l_{0}^{2}w^{D -2}}$ for $D \neq 2$ and $\phi  \sim \frac{GM}{l_{0}^{2}}\ln \left (w\right )$ for $D =2$, is the same in both models, but the detailed expressions are different. This is probably due to the different choices for the differential operators in fractional dimensional spaces used in this work and in Ref. \cite{2020PhRvD.101l4029G}, such as Giusti's fractional Laplacian $\left ( -\Delta \right )^{s}$ as opposed to our choice for the fractional Laplacian in Eq. (\ref{eq3.12}), which will be also studied in detail in the rest of this section.

We note again that both Giusti's model and our NFDG are MOND-like theories, reproducing the asymptotic behavior of MOND, while dropping all the non-linearities, since they are based on differential operators which are actually linear. Both models also introduce similar scale lengths $\ell $ and $l_{0}$, based on the original idea of a MOND\ radius $r_{m} =\sqrt{\frac{GM}{a_{0}}}$ (see footnote 6 in Sect. \ref{sect::gauss}). This specific form of the length scale of these effects can also be justified in terms of a corpuscular interpretation of MOND, as discussed in \cite{2020PhRvD.101l4029G} and references therein.

As also mentioned in the same Ref.  \cite{2020PhRvD.101l4029G}, the proper way to fully describe the transition between the two asymptotic regimes of MOND should be to introduce proper variable-order fractional differential equations, while in this work we modeled the non-linearities of MOND by using our variable dimension $D\left (w\right )$, which is assumed to change slowly over galactic distances so that the fundamental NFDG Eq. (\ref{eq3.11}) is still approximately valid when we replace a fixed dimension $D$ with a variable $D\left (w\right )$.

To conclude this preliminary discussion, we want to mention a few points of contact between NFDG and Verlinde's emergent gravity \cite{2017ScPP....2...16V}. Verlinde considers the entropy content of a de Sitter space for a static coordinate patch described by the metric $ds^{2} = -f\left (r\right )dt^{2} +\frac{dr^{2}}{f\left (r\right )} +r^{2}d\Omega $, where the function $f$ becomes $f\left (r\right ) =1 -\frac{r^{2}}{L^{2}} +2\Phi \left (r\right )$ when a mass $M$ is added at the center. $L$ represents the Hubble scale, while the added mass $M$ is related to the Newtonian potential of the form $\Phi \left (r\right ) = -\frac{8\pi GM}{\left (d -2\right )\Omega _{d -2}r^{d -3}}$. $\Omega _{d -2}$ is the volume of a $\left (d -2\right )$-dimensional unit sphere, where $d$ is the space-time dimension, and $V\left (r\right ) =\Omega _{d -2}r^{d -1}/\left (d -1\right )$ is the volume of a ball of radius $r$ centered at the origin.

Using $d =D +1$ and
$\Omega _{d -2} =\Omega _{D -1} =\int _{S}d\Omega _{D} =\frac{2\pi ^{D/2}}{\Gamma (D/2)} =\frac{2\pi ^{\genfrac{(}{)}{}{}{d -1}{2}}}{\Gamma \genfrac{(}{)}{}{}{d -1}{2}}$, which follows from the dimensional regularization arguments presented in Sect. \ref{sect::gauss}, we can rewrite the above potential as $\Phi \left (r\right ) = -\frac{4\pi ^{1 -\frac{d -1}{2}}\Gamma \genfrac{(}{)}{}{}{d -1}{2}GM}{\left (d -2\right )r^{d -3}}$, which reduces to standard Newtonian for $d =4$. Rewriting this potential in terms of the space dimension $D$, and of the NFDG rescaled coordinates, yields $\phi \left (w\right ) = -\genfrac{[}{]}{}{}{4\pi ^{1 -D/2}\Gamma \left (\frac{D}{2}\right )}{\left (D -1\right )}\frac{GM}{l_{0}^{2}w^{D -2}}$, which shows some similarities with the NFDG and Giusti's potentials presented in the previous paragraphs of this section.

However, Verlinde's theory is not a modification of either the law of inertia or the law of gravity, but rather \cite{2017ScPP....2...16V} ``...the consequence of the emergent nature of gravity... caused by an elastic response due to the volume law contribution to the entanglement entropy in our universe.'' Verlinde derives the MOND phenomenology and the Tully-Fisher relation by setting a fixed $d =4$ space-time dimension, and by using an analogy with the theory of elasticity. Therefore, we don't see any meaningful connection between EG and NFDG.

Going back to the main topic in this section, the Laplace equation in spherical coordinates for a
$D$-dimensional space follows from the
Laplacian operator in Eq. (\ref{eq3.12}):\protect\footnote{ Again, this equation assumes a constant value for the dimension
$D$, although not necessarily an integer one. The theory of NFDG outlined in Sect. \ref{sect::gauss} considered a variable dimension
$D$
as a function of the field point coordinates. If we assume that
$D$
varies slowly with these coordinates, then the results obtained in this section will
be approximately true and valid also in the case of a variable dimension.
}\begin{equation} \nabla _{D}^{2}\phi  =\frac{1}{r^{D -1}}\frac{ \partial }{ \partial r}\left (r^{D -1}\frac{ \partial \phi }{ \partial r}\right ) +\frac{1}{r^{2}\sin ^{D -2}\theta }\frac{ \partial }{ \partial \theta }\left (\sin ^{D -2}\theta \frac{ \partial \phi }{ \partial \theta }\right ) +\frac{1}{r^{2}\sin ^{2}\theta \sin ^{D -3}\varphi }\frac{ \partial }{ \partial \varphi }\left (\sin ^{D -3}\varphi \frac{ \partial \phi }{ \partial \varphi }\right ) =0 , \label{eq6.1}
\end{equation}
where the gravitational potential
$\phi $
can be a function of the standard ``physical'' coordinates
$(r ,\theta  ,\varphi )\text{}$
or of the equivalent dimensionless coordinates
$(w_{r} =r/l_{0} ,w_{\theta } =\theta  ,w_{\varphi } =\varphi )$. In this section, we will simply leave all equations in standard spherical coordinates and
we will omit the scale length
$l_{0}$
in most equations.

 Eq. (\ref{eq6.1}) is separable \cite{doi:10.1063/1.523395,Palmer_2004}, i.e.,
$\phi \left (r ,\theta  ,\varphi \right ) =R(r)\Theta (\theta )\Phi (\varphi )$
and the resulting radial and angular equations are:
\begin{gather}\left [\frac{1}{r^{D -1}}\frac{d}{dr}\left (r^{D -1}\frac{d}{dr}\right ) +\frac{k_{1}}{r^{2}}\right ]R\left (r\right ) =0 \label{eq6.2} \\
\left [\frac{1}{\sin ^{D -2}\theta }\frac{d}{d\theta }\left (\sin ^{D -2}\theta \frac{d}{d\theta }\right ) -k_{1} -\frac{k_{2}}{\sin ^{2}\theta }\right ]\Theta \left (\theta \right ) =0 \nonumber  \\
\left [\frac{1}{\sin ^{D -3}\varphi }\frac{d}{d\varphi }\left (\sin ^{D -3}\varphi \frac{d}{d\varphi }\right ) +k_{2}\right ]\Phi \left (\varphi \right ) =0 \nonumber \end{gather}
where the separation constants take the values
$k_{2} =m\left (m +D -3\right )$,
$m =0 ,1 ,2 , . . .$
and
$k_{1} = -l\left (l +D -2\right )$,
$l =0 ,1 ,2 , . . .$ with
$m \leq l$.

 The complete solution to the angular equations can be found in Appendix B of Ref.
\cite{Palmer_2004}, where both angular functions
$\Theta $
and
$\Phi $
can be expressed in terms of Gegenbauer polynomials \cite{NIST:DLMF,2007tisp.book.....G,morse1953methods}. Here, we will limit our
analysis to cases of axial symmetry, i.e.,
$\phi \left (r ,\theta \right ) =R(r)\Theta (\theta )$
corresponding to
$k_{2} =m =0$
($\Phi (\varphi ) =1$). In this case, the radial equation admits two independent solutions
$R(r) =r^{l}$
and
$R(r) =1/r^{l +D -2}$, while the angular solution is given in terms of Gegenbauer polynomials in
$\cos \theta $:
$\ensuremath{\operatorname*{}}\Theta (\theta ) =C_{l}^{\left (\frac{D}{2} -1\right )}(\cos \theta )$.

Gegenbauer polynomials (a.k.a., ultraspherical functions)
$C_{l}^{\left (\lambda \right )}(x)$
\cite{NIST:DLMF} form a set of orthogonal polynomials defined over the interval
$( -1 ,1)$
and with constraints
$\lambda  > -\frac{1}{2} ,\lambda  \neq 0$. The weight function is
$w(x) =(1 -x^{2})^{\lambda  -\frac{1}{2}}$, i.e., the orthogonality condition is
$\int _{ -1}^{ +1}w(x)C_{l}^{(\lambda )}(x)C_{l^{ \prime }}^{(\lambda )}(x)dx =0$
for
$l \neq l^{ \prime }$
and the normalization factor is
$h_{l} =\frac{2^{1 -2\lambda }\pi \Gamma \left (l +2\lambda \right )}{\left (l +\lambda \right )\left (\Gamma \left (\lambda \right )\right )^{2}l !}$, that is,
$\int _{ -1}^{ +1}\left (C_{l}^{\left (\lambda \right )}(x)\right )^{2}w(x)dx =h_{l}$. The Rodriguez formula for the Gegenbauer polynomials is
$C_{l}^{\left (\lambda \right )}(x) =\frac{1}{\kappa _{l}w(x)}\frac{d^{l}}{dx^{l}}\left (w(x)\left (F(x)\right )^{l}\right )$
with
$F(x) =1 -x^{2}$
and
$\kappa _{l} =\frac{\left ( -2\right )^{l}\left (\lambda  +\frac{1}{2}\right )_{l}l !}{\left (2\lambda \right )_{l}}$,\protect\footnote{ The Pochhammer's symbol
$\left (a\right )_{l}$
is defined as
$\left (a\right )_{0} =1$,
$\left (a\right )_{l} =a\left (a +1\right )\left (a +2\right ) . . .\left (a +l -1\right ) =\Gamma \left (a +l\right )/\Gamma \left (a\right )$.
} and the main generating function is:
$\left (1 -2xz +z^{2}\right )^{ -\lambda } =\sum \limits _{l =0}^{\infty }C_{l}^{\left (\lambda \right )}(x)z^{l}$
for
$\left \vert z\right \vert  <1$.

 Adapting these definitions and properties to the case of our physical solutions
above, we set
$x \equiv \cos \theta $
and
$\lambda  \equiv \frac{D}{2} -1$. The constraints for
$\lambda $
limit the use of Gegenbauer polynomials to the case
$D >1$,
$D \neq 2$
and the overall ortho-normality condition can be written as:

\begin{gather}{\displaystyle\int _{0}^{\pi }}C_{l}^{\left (\frac{D}{2} -1\right )}\left (\cos \theta \right )C_{l^{ \prime }}^{\left (\frac{D}{2} -1\right )}\left (\cos \theta ^{}\right )\sin ^{D -2}\theta d\theta  =h_{l}\delta _{ll^{ \prime }} \label{eq6.3} \\
h_{l} =\frac{2^{3 -D}\pi \Gamma \left (l +D -2\right )}{\left (l +\frac{D}{2} -1\right )\left [\Gamma \left (\frac{D}{2} -1\right )\right ]^{2}l !} . \nonumber \end{gather}
The first few Gegenbauer polynomials in
$\cos \theta $
are:

\begin{equation}C_{0}^{\left (\frac{D}{2} -1\right )}\left (\cos \theta \right ) =1;\ C_{1}^{\left (\frac{D}{2} -1\right )}\left (\cos \theta \right ) =\left (D -2\right )\cos \theta ;\ C_{2}^{\left (\frac{D}{2} -1\right )}\left (\cos \theta \right ) =\left (\frac{D}{2} -1\right )\left (D\cos ^{2}\theta  -1\right ); . . . \label{eq6.4}
\end{equation}

It is evident that, in the case
$D =3$, the Gegenbauer polynomials will reduce to standard Legendre polynomials
$P_{l}\left (\cos \theta \right )$. In the case of possible dependence of the gravitational potential on both angular
variables
$\theta $
and
$\varphi $ \cite{Palmer_2004}, the
ultraspherical functions will reduce instead to standard spherical harmonics
$Y_{l ,m}\left (\theta  ,\varphi \right )$, for
$D =3$. Considering just the case of axial symmetry, the general solution for a given boundary
condition
$\phi \left (R ,\theta \right ) =\phi _{0}\left (\theta \right )$
is:

\begin{gather}\phi (r ,\theta ) =\sum \limits _{l =0}^{\infty }\left (A_{l}r^{l} +\frac{B_{l}}{r^{l +D -2}}\right )C_{l}^{\left (\frac{D}{2} -1\right )}(\cos \theta ) \label{eq6.5} \\
A_{l} =\frac{1}{R^{l}h_{l}}{\displaystyle\int _{0}^{\pi }}\phi _{0}(\theta )C_{l}^{\left (\frac{D}{2} -1\right )}(\cos \theta )\sin ^{D -2}\theta d\theta  \nonumber  \\
B_{l} =\frac{R^{l +D -2}}{h_{l}}{\displaystyle\int _{0}^{\pi }}\phi _{0}\left (\theta \right )C_{l}^{\left (\frac{D}{2} -1\right )}\left (\cos \theta \right )\sin ^{D -2}\theta d\theta  , \nonumber \end{gather}
where the coefficients
$A_{l}$
and
$B_{l}$
were obtained by using Eq. (\ref{eq6.3}).

For a single point-mass placed at the origin,
$\widetilde{\rho }(\mathbf{r}) =\widetilde{m}_{\left (D\right )}\delta ^{\left (D\right )}\left (\mathbf{r}\right )$, the potential following the first line of Eq. (\ref{eq6.5}) must consist only of the
$l =0$
term, due to spherical symmetry. Requiring also the potential to be zero at infinity
implies
$\phi \left (r\right ) =\frac{B_{0}}{r^{D -2}}$, from which
$ \nabla _{D}^{2}\phi  = -\frac{B_{0}\left (D -2\right )2\pi ^{D/2}\delta ^{\left (D\right )}\left (\mathbf{r}\right )}{\Gamma \left (D/2\right )}$, following
$ \nabla _{D} \cdot \genfrac{(}{)}{}{}{\widehat{\mathbf{r}}}{r^{D -1}} =\frac{2\pi ^{D/2}}{\Gamma \left (D/2\right )}\delta ^{(D)}(\mathbf{r})$
and
$ \nabla _{D}\genfrac{(}{)}{}{}{1}{r^{D -2}} = -\left (D -2\right )\frac{\widehat{\mathbf{r}}}{r^{D -1}}$. On the other hand, the generalized Poisson equation (\ref{eq3.16}) implies
$ \nabla _{D}^{2}\phi  =4\pi G\widetilde{m}_{\left (D\right )}\delta ^{\left (D\right )}\left (\mathbf{r}\right )$
and combining both results we obtain the coefficient
$B_{0} = -\frac{2\pi ^{1 -D/2}\Gamma \left (D/2\right )G\widetilde{m}_{\left (D\right )}}{\left (D -2\right )}$
and the potential of a point-mass becomes
$\phi \left (r\right ) = -\frac{2\pi ^{1 -D/2}\Gamma \left (D/2\right )G\widetilde{m}_{\left (D\right )}}{\left (D -2\right )r^{D -2}}$, thus confirming the first line of Eq. (\ref{eq3.11}).

Since the fractal gradient we introduced in Sect. \ref{sect::gauss} is the same as the standard one, the gravitational field
we obtain from the point mass potential above, confirms also our previous Eq. (\ref{eq3.5}) in Sect. \ref{sect::gauss}
which was originally introduced in a heuristic way. In fact, adding the appropriate scale length
$l_{0}$
into the equations for dimensional correctness, we obtain
$\mathbf{g} = - \nabla _{D}\phi  = -\frac{2\pi ^{1 -D/2}\Gamma \left (D/2\right )G\widetilde{m}_{\left (D)\right .}}{l_{0}^{2}}\frac{\left (\mathbf{r}/l_{0}\right )}{\left (r/l_{0}\right )^{D}}$, which is equivalent to Eq. (\ref{eq3.5}).

By using the generating function of the Gegenbauer polynomials above,
rewritten with
$x =\cos \theta ^{ \prime }$,
$\lambda  =\frac{D}{2} -1$, and
$z =r^{ \prime }/r$
or
$z =r/r^{ \prime }$, for
$r^{ \prime } <r$
and
$r <r^{ \prime }$
respectively, we can write:

\begin{equation}\frac{1}{\left \vert \mathbf{r} -\mathbf{r}^{ \prime }\right \vert ^{D -2}} =\left \{\begin{array}{c}\frac{1}{r^{D -2}}\sum \limits _{l =0}^{\infty }\genfrac{(}{)}{}{}{r^{ \prime }}{r}^{l}C_{l}^{\left (\frac{D}{2} -1\right )}\left (\cos \theta ^{ \prime }\right );\text{  fo}\text{r }r^{ \prime } <r\text{
} \\
\frac{1}{r^{ \prime D -2}}\sum \limits _{l =0}^{\infty }\genfrac{(}{)}{}{}{r}{r^{ \prime }}^{l}C_{l}^{\left (\frac{D}{2} -1\right )}\left (\cos \theta ^{ \prime }\right );\text{  for }r <r^{ \prime }\text{
}\end{array}.\right . \label{eq6.6}
\end{equation}

Using the first line of the last equation combined with Eq. (\ref{eq3.11}), we can derive a generalized multipole expansion for a localized
charge distribution
$\rho  =\rho (r^{ \prime } ,\theta ^{ \prime })$
in the case
$r \gg r^{ \prime }$:

\begin{equation}\phi \left (\mathbf{r}\right ) = -\frac{2\pi ^{1 -D/2}\Gamma \left (D/2\right )G}{\left (D -2\right )}\int _{V_{D}}\frac{\widetilde{\rho }\left (\mathbf{r}^{ \prime }\right )dV_{D}}{\left \vert r -r^{ \prime }\right \vert ^{D -2}} = -\frac{2\pi ^{1 -D/2}\Gamma \left (D/2\right )G}{\left (D -2\right )}{\displaystyle\sum _{l =0}^{\infty }}\frac{1}{r^{D -2 +l}}{\displaystyle\int _{V_{D}}}\widetilde{\rho }(r^{ \prime })\left (r^{ \prime }\right )^{l}C_{l}^{\left (\frac{D}{2} -1\right )}\left (\cos \theta ^{ \prime }\right )dV_{D} , \label{eq6.7}
\end{equation}
with the first two terms representing the monopole and dipole moments:

\begin{gather}\phi _{mon}\left (\mathbf{r}\right ) = -\frac{2\pi ^{1 -D/2}\Gamma \left (D/2\right )G}{\left (D -2\right )r^{D -2}}\int _{V_{D}}\widetilde{\rho }\left (\mathbf{r}^{ \prime }\right )dV_{D} = -\frac{2\pi ^{1 -D/2}\Gamma \left (D/2\right )G\widetilde{M}_{(D)}}{(D -2)r^{D -2}} \label{eq6.8} \\
\phi _{dip}\left (\mathbf{r}\right ) = -\frac{2\pi ^{1 -D/2}\Gamma \left (D/2\right )G}{\left (D -2\right )r^{D -1}}\int _{V_{D}}\widetilde{\rho }\left (\mathbf{r}^{ \prime }\right )r^{ \prime }\left (D -2\right )\cos \theta ^{ \prime }dV_{D} = -\frac{2\pi ^{1 -D/2}\Gamma \left (D/2\right )G\widetilde{\mathbf{p}}_{(D)} \cdot \widehat{\mathbf{r}}}{r^{D -1}} , \nonumber \end{gather}
where
$\widetilde{M}_{\left (D\right )}$
is the total D-dimensional mass and
$\widetilde{\mathbf{p}}_{\left (D\right )} =\int _{V_{D}}\mathbf{r}^{ \prime }\rho \left (\mathbf{r}^{ \prime }\right )dV_{D}$
is the D-dimensional dipole moment.

 Again, using the standard gradient operator it is easy to obtain the corresponding
monopole and dipole terms for the gravitational field:

\begin{gather}\mathbf{g}_{mon}\left (\mathbf{r}\right ) = -\frac{2\pi ^{1 -D/2}\Gamma \left (D/2\right )G\widetilde{M}_{(D)}}{r^{D -1}}\widehat{\mathbf{r}} \label{eq6.9} \\
\mathbf{g}_{dip}\left (\mathbf{r}\right ) = -\frac{2\pi ^{1 -D/2}\Gamma \left (D/2\right )G\widetilde{\mathbf{p}}_{\left (D\right )}}{r^{D}}\left [\left (D -1\right )\cos \theta \ \widehat{\mathbf{r}} +\sin \theta \ \widehat{\boldsymbol{\theta }}\right ] , \nonumber \end{gather}
with the monopole term consistent with our previous analysis and with Eq. (\ref{eq3.5}) in Sect. \ref{sect::gauss}.

We can also consider the potential of a thin shell of radius
$a$
and surface mass density
$\sigma \left (\theta \right )$
at fixed radius
$r =a$. In this case ``interior'' ($r \leq a$) and ``exterior'' ($r \geq a$) solutions can be written as:

\begin{gather}\phi _{int}\left (r ,\theta \right ) =\sum \limits _{l =0}^{\infty }A_{l}r^{l}C_{l}^{\left (\frac{D}{2} -1\right )}\left (\cos \theta \right ) \label{eq6.10} \\
\phi _{ext}\left (r ,\theta \right ) =\sum \limits _{l =0}^{\infty }\frac{B_{l}}{r^{l +D -2}}C_{l}^{\left (\frac{D}{2} -1\right )}\left (\cos \theta \right ) \nonumber \end{gather}
since the potential at the center must be non singular and the potential at infinity is
assumed to be zero. Continuity at
$r =a$
and orthogonality of the
$C_{l}$
functions also implies
$B_{l} =A_{l}a^{2l +D -2}$.

 Following a procedure similar to the one outlined in Sect. 2.4 of Ref. \cite{2008gady.book.....B} for the standard
$D =3$
case, we can expand the surface mass density of the thin shell as
$\sigma (\theta ) =\sum \limits _{l =0}^{\infty }\sigma _{l}C_{l}^{\left (\frac{D}{2} -1\right )}\left (\cos \theta \right )$
and, using the ortho-normality of the
$C_{l}$
functions, also obtain:

\begin{equation}\sigma _{l} =\frac{1}{h_{l}}\int _{0}^{\pi }\sigma \left (\theta \right )C_{l}^{\left (\frac{D}{2} -1\right )}\left (\cos \theta \right )\sin ^{D -2}\theta d\theta  . \label{eq6.11}
\end{equation}

Gauss's theorem (\ref{eq3.14}) applied to a
small piece of the shell yields
$\genfrac{[}{]}{}{}{ \partial \phi _{ext}}{ \partial r}_{r =a} -\genfrac{[}{]}{}{}{ \partial \phi _{int}}{ \partial r}_{r =a} =4\pi G\sigma \left (\theta \right )$
so that combining this equation with Eq. (\ref{eq6.10})
and the
$\sigma \left (\theta \right )$
expansion above, another relation between the
$A_{l}$
and the
$B_{l}$
coefficients can be derived:
$ -\sum \limits _{l =0}^{\infty }\left [\left (l +D -2\right )B_{l}a^{ -\left (l +D -1\right )} +lA_{l}a^{l -1}\right ]C_{l}^{\left (\frac{D}{2} -1\right )}\left (\cos \theta \right ) =4\pi G\sum \limits _{l =0}^{\infty }\sigma _{l}C_{l}^{\left (\frac{D}{2} -1\right )}\left (\cos \theta \right )$. Using this relation and the previous one between these coefficients, it is then possible
to extract them explicitly:

\begin{gather}A_{l} = -\frac{4\pi Ga^{ -\left (l -1\right )}}{2l +D -2}\sigma _{l} \label{eq6.12} \\
B_{l} = -\frac{4\pi Ga^{l +D -1}}{2l +D -2}\sigma _{l} \nonumber \end{gather}

 The interior and exterior potentials are then obtained:

\begin{gather}\phi _{int}\left (r ,\theta \right ) = -4\pi Ga\sum \limits _{l =0}^{\infty }\genfrac{(}{)}{}{}{r}{a}^{l}\frac{\sigma _{l}}{2l +D -2}C_{l}^{\left (\frac{D}{2} -1\right )}\left (\cos \theta \right ) \label{eq6.13} \\
\phi _{ext}\left (r ,\theta \right ) = -4\pi Ga\sum \limits _{l =0}^{\infty }\genfrac{(}{)}{}{}{a}{r}^{l +D -2}\frac{\sigma _{l}}{2l +D -2}C_{l}^{\left (\frac{D}{2} -1\right )}\left (\cos \theta \right ) \nonumber \end{gather}
with
$\sigma _{l}$
given by Eq. (\ref{eq6.11}). Finally, the potential of
a solid body is evaluated by breaking it into a series of concentric shells. Let
$\delta \sigma _{l}\left (a\right )$
be the
$\sigma $-coefficient of the shell lying between
$a$
and
$a +\delta a$, and
$\delta \phi \left (r ,\theta  ;a\right )$
the corresponding potential at
$r$. Following Eq. (\ref{eq6.11}) with
$\sigma \left (\theta \right ) =\rho \left (a ,\theta \right )\delta a$
we can write
$\delta \sigma _{l}\left (a\right ) =\left [\frac{1}{h_{l}}\int _{0}^{\pi }d\theta \sin ^{D -2}\theta C_{l}^{\left (\frac{D}{2} -1\right )}\left (\cos \theta \right )\rho \left (a ,\theta \right )\right ]\delta a \equiv \rho _{l}\left (a\right )\delta a$. Substituting these values of
$\delta \sigma _{l}$
into Eq. (\ref{eq6.13}) and summing all contributions
over all possible values for
$a$
we obtain the potential at
$r$
generated by the entire collection of shells:

\begin{gather}\phi \left (r^{} ,\theta \right ) =\sum \limits _{a =0}^{r}\delta \phi _{ext} +\sum \limits _{a =r}^{\infty }\delta \phi _{int} \label{eq6.14} \\
 = -4\pi G\sum \limits _{l =0}^{\infty }\frac{C_{l}^{\left (\frac{D}{2} -1\right )}\left (\cos \theta \right )}{2l +D -2}\left [\frac{1}{r^{l +D -2}}\int _{0}^{r}da\thinspace a^{l +D -1}\rho _{l}\left (a\right ) +r^{l}\int _{r}^{\infty }\frac{da}{a^{l -1}}\rho _{l}\left (a\right )\right ] \nonumber  \\
\rho _{l}\left (a\right ) =\frac{1}{h_{l}}\int _{0}^{\pi }d\theta \sin ^{D -2}\theta \thinspace C_{l}^{\left (\frac{D}{2} -1\right )}\left (\cos \theta \right )\rho \left (a ,\theta \right ) , \nonumber \end{gather}
with
$h_{l}$
given by Eq. (\ref{eq6.3}).

 Eq. (\ref{eq6.14}) represents the multipole
expansion for the gravitational potential
$\phi $
in the case of an axially symmetric distribution of matter, for a generalized dimension
$D >1$
($D \neq 2$). A more general expansion formula can be derived for the case
$\phi \left (r ,\theta  ,\varphi \right )$
by using two sets of ultraspherical functions, but this goes beyond the scope of the
current work. It is easy to check that the multipole expansion above for the simple case of a point
mass at the origin, i.e.,
$\widetilde{\rho }\left (a\right ) =\widetilde{m}_{\left (D\right )}\delta ^{\left (D\right )}\left (\mathbf{a}\right ) =\widetilde{m}_{\left (D\right )}\frac{\Gamma \left (D/2\right )}{2\pi ^{D/2}a^{D -1}}\delta \left (a\right )$, would be limited to the
$l =0$
terms and yield
$\phi \left (r ,\theta \right ) = -4\pi G\frac{1}{D -2}\left [\frac{1}{r^{D -2}}\int _{0}^{r}daa^{D -1}\widetilde{m}_{\left (D\right )}\frac{\Gamma \left (D/2\right )}{2\pi ^{D/2}a^{D -1}}\delta \left (a\right )\right ] = -\frac{2\pi ^{1 -D/2}\Gamma \left (D/2\right )G\widetilde{m}_{\left (D\right )}}{\left (D -2\right )r^{D -2}}$, which again confirms our previous equation (\ref{eq3.11}), for
$D \neq 2$.

The special case
$D =2$
corresponds to the classic Dirichlet problem for a circle and the Poisson kernel (see for
example Ref. \cite{smirnov2014course}). In fact, adapting Eqs. (\ref{eq6.1}) and (\ref{eq6.2}) to
the case
$D =2$
and considering a separable solution of the type
$\phi \left (r ,\theta \right ) =R(r)\Theta (\theta )$, as was done for the general
$D \neq 2$
case above, we obtain:

\begin{gather} \nabla _{(2)}^{2}\phi  =\frac{1}{r}\frac{ \partial }{ \partial r}\left (r\frac{ \partial \phi }{ \partial r}\right ) +\frac{1}{r^{2}}\frac{ \partial ^{2}\phi }{ \partial \theta ^{2}} =0 \label{eq6.15} \\
\left [\frac{1}{r}\frac{d}{dr}\left (r\frac{d}{dr}\right ) -\frac{l^{2}}{r^{2}}\right ]R\left (r\right ) =0 \nonumber  \\
\left [\frac{d^{2}}{d\theta ^{2}} +l^{2}\right ]\Theta \left (\theta \right ) =0 , \nonumber \end{gather}
where
$\theta $
can now be considered the polar angle and
$l =0 ,1 ,2 , . . .$
as before. Solutions to the radial and angular equations are:

\begin{gather}\phi _{0}\left (r ,\theta \right ) =(A_{0} +B_{0}\theta )\ (C_{0}\ln r +D_{0})_{} \label{eq6.16} \\
\phi _{l}\left (r ,\theta \right ) =[A_{l}\cos \left (l\theta \right ) +B_{l}\sin \left (l\theta \right )]\ (C_{l}r^{l} +D_{l}r^{ -l}\ ) ;\ l \neq 0 \nonumber \end{gather}

 Setting
$B_{0} =0$, due to the periodicity of the solution, renaming the constants
$A_{0}D_{0} \rightarrow \frac{a_{0}}{2} ,\frac{a_{0}^{ \prime }}{2}$,
$A_{0}C_{0} \rightarrow \frac{c_{0}}{2}$,
$A_{l}C_{l} \rightarrow a_{l}$,
$B_{l}C_{l} \rightarrow b_{l}$,
$A_{l}D_{l} \rightarrow c_{l}$, and
$B_{l}D_{l} \rightarrow d_{l}$, we can write the interior and exterior solutions to a circle of radius
$R$
as:\protect\footnote{ We still require the potential at the center to be non singular, but we will
allow the potential at infinity to diverge at most logarithmically.
}

\begin{gather}\phi _{int}\left (r ,\theta \right ) =\frac{a_{0}}{2} +\sum \limits _{l =1}^{\infty }\left [a_{l}\cos \left (l\theta \right ) +b_{l}\sin \left (l\theta \right )\right ]r^{l} \label{eq6.17} \\
\phi _{ext}\left (r ,\theta \right ) =\frac{a_{0}^{ \prime }}{2} +\frac{c_{0}}{2}\ln r +\sum \limits _{l =1}^{\infty }\left [c_{l}\cos \left (l\theta \right ) +d_{l}\sin \left (l\theta \right )\right ]r^{ -l} , \nonumber \end{gather}
respectively for
$0 \leq r \leq R$
and
$r \geq R$. For a given boundary condition on the circle of radius
$R$:
$\phi \left (R ,\theta \right ) \equiv \phi _{0}\left (\theta \right )$, it is easy to obtain all coefficients as follows:

\begin{gather}a_{l} =\frac{1}{\pi R^{l}}\int _{ -\pi }^{\pi }\phi _{0}\left (\theta ^{ \prime }\right )\cos \left (l\theta ^{ \prime }\right )d\theta ^{ \prime } \label{eq6.18} \\
b_{l} =\frac{1}{\pi R^{l}}\int _{ -\pi }^{\pi }\phi _{0}\left (\theta ^{ \prime }\right )\sin \left (l\theta ^{ \prime }\right )d\theta ^{ \prime } \nonumber  \\
c_{l} =\frac{R^{l}}{\pi }\int _{ -\pi }^{\pi }\phi _{0}\left (\theta ^{ \prime }\right )\cos \left (l\theta ^{ \prime }\right )d\theta ^{ \prime } \nonumber  \\
d_{l} =\frac{R^{l}}{\pi }\int _{ -\pi }^{\pi }\phi _{0}\left (\theta ^{ \prime }\right )\sin \left (l\theta ^{ \prime }\right )d\theta ^{ \prime } \nonumber \end{gather}
where these also include the
$a_{0}$
and
$c_{0}$
coefficients in Eq. (\ref{eq6.17}) ($b_{0}$
and
$d_{0}$
are identically zero). The
$a_{0}^{ \prime }$
coefficient is obtained instead by equating the
$l =0$
interior and exterior solutions in Eq. (\ref{eq6.17})
at the boundary
$r =R$, namely
$a_{0}^{ \prime } +c_{0}\ln R =a_{0}$, from which
$\phi _{ext}^{l =0}\left (r ,\theta \right ) =\frac{a_{0}}{2} +\frac{c_{0}}{2}\ln \genfrac{(}{)}{}{}{r}{R}$. Inserting all these expressions into Eq. (\ref{eq6.17}) and using the identity
$1 +2\sum \limits _{l =1}^{\infty }x^{l}\cos \left (l\theta \right ) =\frac{1 -x^{2}}{x^{2} -2x\cos \theta  +1}$
($0 \leq x <1$) we can rewrite both interior and exterior solutions as power series or with Poisson
integrals:

\begin{gather}\phi _{int}\left (r ,\theta \right ) =\frac{1}{2\pi }\int _{ -\pi }^{\pi }\phi _{0}\left (\theta ^{ \prime }\right )d\theta ^{ \prime } +\frac{1}{\pi }\sum \limits _{l =1}^{\infty }\left [\int _{ -\pi }^{\pi }\phi _{0}\left (\theta ^{ \prime }\right )\cos \left (l\left (\theta ^{ \prime } -\theta \right )\right )d\theta ^{ \prime }\right ]\genfrac{(}{)}{}{}{r}{R}^{l} \label{eq6.19} \\
 =\frac{1}{2\pi }\int _{ -\pi }^{\pi }\phi _{0}\left (\theta ^{ \prime }\right )\frac{1 -\genfrac{(}{)}{}{}{r}{R}^{2}}{\genfrac{(}{)}{}{}{r}{R}^{2} -2\genfrac{(}{)}{}{}{r}{R}\cos \left (\theta ^{ \prime } -\theta \right ) +1}d\theta ^{ \prime } \nonumber  \\
\phi _{ext}\left (r ,\theta \right ) =\frac{1}{2\pi }\int _{ -\pi }^{\pi }\phi _{0}\left (\theta ^{ \prime }\right )d\theta ^{ \prime }\left (1 +\ln \genfrac{(}{)}{}{}{r}{R}\right ) +\frac{1}{\pi }\sum \limits _{l =1}^{\infty }\left [\int _{ -\pi }^{\pi }\phi _{0}\left (\theta ^{ \prime }\right )\cos \left (l\left (\theta ^{ \prime } -\theta \right )\right )d\theta ^{ \prime }\right ]\genfrac{(}{)}{}{}{R}{r}^{l} \nonumber  \\
 =\frac{1}{2\pi }\int _{ -\pi }^{\pi }\phi _{0}\left (\theta ^{ \prime }\right )\left [\ln \genfrac{(}{)}{}{}{r}{R} +\frac{1 -\genfrac{(}{)}{}{}{R}{r}^{2}}{\genfrac{(}{)}{}{}{R}{r}^{2} -2\genfrac{(}{)}{}{}{R}{r}\cos \left (\theta ^{ \prime } -\theta \right ) +1}\right ]d\theta ^{ \prime } . \nonumber \end{gather}

 As it was done for the case
$D \neq 2$, we can consider a single point mass at the origin in a
$D =2$
space as
$\widetilde{\rho }\left (\mathbf{r}\right ) =\widetilde{m}_{\left (2\right )}\delta ^{\left (2\right )}\left (\mathbf{r}\right )$. The spherical symmetry of
$\phi _{ext}\left (r\right )$
obtained from the third line of Eq. (\ref{eq6.19})
requires
$l =0$ and neglecting constant terms the potential simplifies as
$\phi \left (r\right ) =C\ln r$, where
$C$
is some constant to be determined. Following the procedure used for the
$D \neq 2$
above, a direct computation gives
$ \nabla _{\left (2\right )}^{2}\phi \left (r\right ) =2\pi C\delta ^{\left (2\right )}\left (\mathbf{r}\right )$
while the generalized Poisson equation (\ref{eq3.16})
implies
$ \nabla _{\left (2\right )}^{2}\phi \left (r\right ) =4\pi G\widetilde{m}_{\left (2\right )}\delta ^{\left (2\right )}\left (\mathbf{r}\right )$. Comparing the two results, we determine the constant
$C =2G\widetilde{m}_{\left (2\right )}$
and
$\phi \left (r\right ) =2G\widetilde{m}_{\left (2\right )}\ln r \rightarrow \frac{2G\widetilde{m}_{\left (2\right )}}{l_{0}^{2}}\ln \left (r/l_{0}\right )$, including also the scale length
$l_{0}$
for dimensional correctness. This result confirms the second line of our general equation
(\ref{eq3.11}) for the gravitational potential in the
$D =2$
case.

\section{\label{sect:appendix_spherical}
Spherically symmetric mass distributions
}
 In this section we will consider the gravitational field
$\mathbf{g}(w)$
due to a spherically symmetric source mass distribution
$\widetilde{\rho }(w^{ \prime })$, in a fractal space of dimension $D\left (w\right )$ also depending on the distance from the center of the coordinate system. We will expand and
generalize the standard Newtonian derivation, based on the computation of the field due to an
infinitesimal spherical shell (see for example \cite{fowles2005analytical}, pp 223-225). Since the gravitational field can
be computed directly from the mass distribution, following Eq. (\ref{eq3.8}), we will not use the gravitational potential
$\phi $
in this section.

\begin{figure}\centering 
\setlength\fboxrule{0in}\setlength\fboxsep{0.1in}\fcolorbox[HTML]{000000}{FFFFFF}{\includegraphics[ width=2.532171in, height=1.6742770000000002in,]{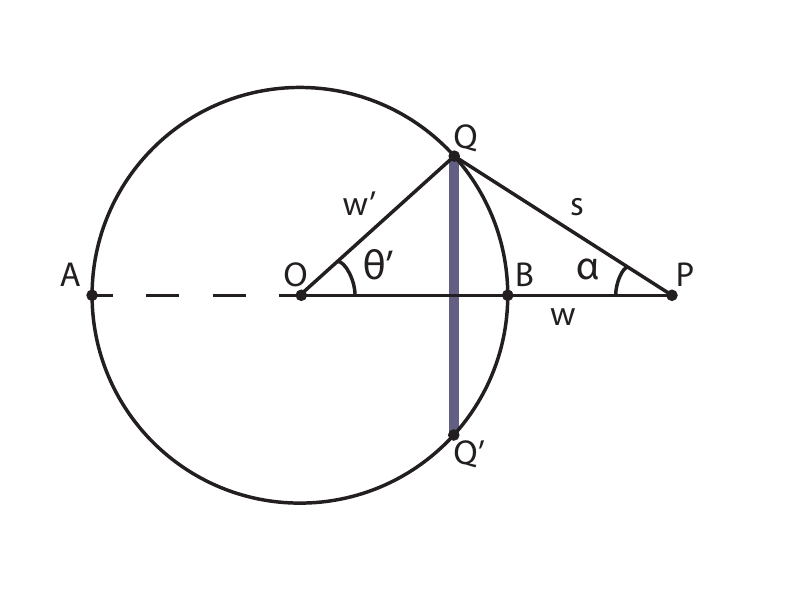}
}
\caption{General geometry used for spherical mass
distributions.
}\label{figure: Geometry}\end{figure}

 The geometry is illustrated in Fig. \ref{figure: Geometry}. The distance
$s$
between the source point
$Q$
and the field point
$P$
is equivalent to the quantity$\left .\right .$$\left \vert \mathbf{w} -\mathbf{w}^{ \prime }\right \vert $
in Eq. (\ref{eq3.8}), with
$w^{2} +w^{ \prime 2} -2ww^{ \prime }\cos \theta ^{ \prime } =s^{2}$
due to the law of cosines and, by differentiating,
$ww^{ \prime }\sin \theta ^{ \prime }d\theta ^{ \prime } =sds$. We also note that
$\sin \theta ^{ \prime } =\sqrt{1 -\cos ^{2}\theta ^{ \prime }} =\sqrt{1 -\frac{\left (w^{2} +w^{ \prime 2} -s^{2}\right )^{2}}{4w^{2}w^{ \prime 2}}}$
and that in the triangle
$OPQ$
we also have
$\cos \alpha  =\frac{s^{2} +w^{2} -w^{ \prime 2}}{2ws}$, which is useful to project the field into the radial direction
$OP$.  For a uniform spherical shell of radius
$w^{ \prime }$
and dimension
$2 <D \leq 3$, the mass of the infinitesimal ring outlined in gray in the figure between points
$Q$
and
$Q^{ \prime }$
is:

\begin{gather}d\widetilde{m}_{(D)} =\frac{\pi ^{\frac{D}{2} -1}}{\Gamma \left (\frac{D}{2} -1\right )}\widetilde{\varrho }(w^{ \prime })w^{ \prime D -1}dw^{ \prime }\left \vert \sin \theta ^{ \prime }\right \vert ^{D -2}d\theta ^{ \prime }\int _{0}^{2\pi }\left \vert \cos \varphi ^{ \prime }\right \vert ^{D -3}d\varphi ^{ \prime } \label{eq7.1} \\
 =\frac{2\pi ^{\frac{D -1}{2}}}{\Gamma \genfrac{(}{)}{}{}{D -1}{2}}\widetilde{\varrho }(w^{ \prime })w^{ \prime ^{D -1}}dw^{ \prime }\left \vert \sin \theta ^{ \prime }\right \vert ^{D -2}d\theta ^{ \prime } . \nonumber \end{gather}

The previous equation follows from Eq. (\ref{eq3.9}) expressed in terms of spherical coordinates, as discussed in Sect. \ref{sect::gauss}. Following Eq. (\ref{eq3.8}), we can obtain the infinitesimal contribution to the field at point
$P$, due to the whole spherical shell, by integrating over the angle
$\theta ^{ \prime }$:

\begin{gather}d\mathbf{g}(w) = -\frac{4\sqrt{\pi }G}{l_{0}^{2}}\frac{\Gamma \left (D/2\right )}{\Gamma \genfrac{(}{)}{}{}{D -1}{2}}\widetilde{\rho }(w^{ \prime })w^{ \prime D -1}dw^{ \prime }\int _{0}^{\pi }\left \vert \sin \theta ^{ \prime }\right \vert ^{D -2}\frac{1}{s^{D -1}}\frac{s^{2} +w^{2} -w^{ \prime 2}}{2ws}d\theta ^{ \prime }\widehat{\mathbf{w}} \label{eq7.2} \\
 = -\frac{4\sqrt{\pi }G}{l_{0}^{2}}\frac{\Gamma \left (D/2\right )}{\Gamma \genfrac{(}{)}{}{}{D -1}{2}}\frac{\widetilde{\varrho }(w^{ \prime })w^{ \prime }dw^{ \prime }}{2^{D -2}w^{D -1}}{\displaystyle\int _{w -w^{ \prime }}^{w +w^{ \prime }}}\left [\left (2w^{ \prime }w\right )^{2} -\left (w^{2} +w^{ \prime 2} -s^{2}\right )^{2}\right ]^{\frac{D -3}{2}}\genfrac{(}{)}{}{}{s^{2} +w^{2} -w^{ \prime 2}}{s^{D -1}}ds\ \widehat{\mathbf{w}} \nonumber  \\
 = -\frac{4\pi G}{l_{0}^{2}w^{D -1}}\widetilde{\rho }(w^{ \prime })w^{ \prime D -1}dw^{ \prime }\widehat{\mathbf{w}} . \nonumber \end{gather}

Due to the symmetry, in this equation we considered just the radial components of
the field at point
$P$, while the sum of the perpendicular components vanishes, and we used all the relations
mentioned above to express the various quantities in terms of the variable
$s$. The angular integration over
$\theta ^{ \prime }$
was then transformed into an integration over
$s$, by assuming
$w >w^{ \prime }$
and the integral in the second line of Eq. (\ref{eq7.2})
is:
${\displaystyle\int _{w -w^{ \prime }}^{w +w^{ \prime }}}\left [\left (2w^{ \prime }w\right )^{2} -\left (w^{2} +w^{ \prime 2} -s^{2}\right )^{2}\right ]^{\frac{D -3}{2}}\genfrac{(}{)}{}{}{s^{2} +w^{2} -w^{ \prime 2}}{s^{D -1}}ds =\frac{2^{D -2}\Gamma \left (1 -\frac{D}{2}\right )\Gamma \genfrac{(}{)}{}{}{D -1}{2}\sin \genfrac{(}{)}{}{}{D\pi }{2}w^{ \prime D -2}}{\sqrt{\pi }}$, which yields the simplified result on the third line of the last equation, in terms of the
radial unit vector
$\widehat{\mathbf{w}}$.

 The total field
$\mathbf{g}(w)$
at point
$P$
can then be obtained with a further integration over all the ``inner'' spherical shells ($0 <w^{ \prime } <w$) noting that the dimension
$D$
should also be considered as a function of the field point radius
$w$, i.e.
$D =D(w)$, therefore:

\begin{equation}\mathbf{g}(w) = -\frac{4\pi G}{l_{0}^{2}w^{D\left (w\right ) -1}}{\displaystyle\int _{0}^{w}}\tilde{\rho }\left (w^{ \prime }\right )w^{ \prime ^{D\left (w\right ) -1}}dw^{ \prime }\widehat{\mathbf{w}} . \label{eq7.3}
\end{equation}

The standard Newtonian result
$\mathbf{g}(w) = -\frac{4\pi G}{l_{0}^{2}w^{2}}\int _{0}^{w}\widetilde{\rho }(w^{ \prime })w^{ \prime 2}dw^{ \prime }\widehat{\mathbf{w}}$
is recovered from the previous equation in the case of a constant dimension
$D =3$. We should also remark that, as in the standard Newtonian case, all contributions to the
field
$\mathbf{g}(w)$
due to the ``outer'' spherical shells, that is for
$w^{ \prime } >w >0$, are identically zero also in the case of variable dimension
$D(w)$. In fact, for the outer shells the integral in the second line of Eq. (\ref{eq7.2}) would only differ by the lower limit:
$w^{ \prime } -w$, instead of
$w -w^{ \prime }$. It is easy to check that
${\displaystyle\int _{w^{ \prime } -w}^{w +w^{ \prime }}}\left [\left (2w^{ \prime }w\right )^{2} -\left (w^{2} +w^{ \prime 2} -s^{2}\right )^{2}\right ]^{\frac{D -3}{2}}\genfrac{(}{)}{}{}{s^{2} +w^{2} -w^{ \prime 2}}{s^{D -1}}ds =0$
and, therefore, outer spherical shells do not contribute to the field. Newton's first theorem
for a spherical distribution of matter, namely that
\textit{a body inside a spherical shell of
matter experiences no net gravitational force from that
shell}, still holds. 

 The main result in Eq. (\ref{eq7.3}) was derived
for the case
$2 <D \leq 3$, but it is easy to check that this result is actually valid for the whole range
$1 \leq D \leq 3$. The special case
$D =2$
involves a ``circular'' distribution of matter, over the circle of radius
$w^{ \prime }$ in Fig. 1. In this case, the ``ring'' reduces to the two symmetric points
$Q$
and
$Q^{ \prime }$, with the infinitesimal ``ring'' mass
$d\widetilde{m}_{(2)} =2\widetilde{\rho }(w^{ \prime })w^{ \prime }dw^{ \prime }d\theta ^{ \prime }$, where the factor of 2 accounts for the two equal contributions from points
$Q$
and
$Q^{ \prime }$, while the
$\theta ^{ \prime }$
integration is between
$0$
and
$\pi $. In this way, the derivation is similar to the one outlined above and Eq. (\ref{eq7.3}) still holds. In particular, for a space of constant
$D =2$, we have
$\mathbf{g}(w) = -\frac{4\pi G}{l_{0}^{2}w}\int _{0}^{w}\widetilde{\rho }(w^{ \prime })w^{ \prime }dw^{ \prime }\widehat{\mathbf{w}}$.

 The case
$1 <D <2$
would still be related to a ``circular'' distribution of matter, but with the infinitesimal
mass contribution given by
$d\widetilde{m}_{(D)} =2\frac{\pi ^{\frac{D -1}{2}}}{\Gamma \genfrac{(}{)}{}{}{D -1}{2}}\widetilde{\rho }(w^{ \prime })w^{ \prime D -1}dw^{ \prime }\left \vert \sin \theta ^{ \prime }\right \vert ^{D -2}d\theta ^{ \prime }$, which is equivalent to the result in the second line of Eq. (\ref{eq7.1}). Therefore, the general result in Eq. (\ref{eq7.3}) also follows in this case, since it can be shown that the ``outer'' spherical shells contributions are identically zero also in this case.

For
$0 <D \leq 1$, the ``spherical'' distribution of matter reduces just to points
$A$
and
$B$
in the figure. The infinitesimal mass is $d\widetilde{m}_{(D)} =\frac{\pi ^{D/2}}{\Gamma \left (D/2\right )}\widetilde{\rho }(w^{ \prime })w^{ \prime D -1}dw^{ \prime }$ for each of the two points, which reduces to
$d\widetilde{m}_{(1)} =\widetilde{\rho }(w^{ \prime })dw^{ \prime }$ for
$D =1$. Using Eq. (\ref{eq3.8}), the infinitesimal contribution to the field at        $P$, due to source points
$A$
and
$B$, must now be divided between the inner shells,
$d\mathbf{g}_{inner}(w) =[ -\frac{2\pi G}{l_{0}^{2}}\widetilde{\rho }\left (w^{ \prime }\right )\frac{w^{ \prime D -1}}{\left (w -w^{ \prime }\right )^{D -1}}dw^{ \prime } -\frac{2\pi G}{l_{0}^{2}}\widetilde{\rho }\left (w^{ \prime }\right )\frac{w^{ \prime D -1}}{\left (w +w^{ \prime }\right )^{D -1}}dw^{ \prime }]\widehat{\mathbf{w}}$, and the outer shells, $d\mathbf{g}_{outer}(w) =[ +\frac{2\pi G}{l_{0}^{2}}\widetilde{\rho }\left (w^{ \prime }\right )\frac{w^{ \prime D -1}}{\left (w^{ \prime } -w\right )^{D -1}}dw^{ \prime } -\frac{2\pi G}{l_{0}^{2}}\widetilde{\rho }\left (w^{ \prime }\right )\frac{w^{ \prime D -1}}{\left (w +w^{ \prime }\right )^{D -1}}dw^{ \prime }]\widehat{\mathbf{w}}$.  In general, the
overall field at point
$P$
is due to both inner and outer contributions. Integrating the previous expressions and simplifying, we obtain:

\begin{gather}\mathbf{g}\left (w\right ) =\mathbf{g}_{inner}\left (w\right ) +\mathbf{g}_{outer}\left (w\right ) \label{eq7.4} \\
\mathbf{g}_{inner}(w) = -\frac{2\pi G}{l_{0}^{2}}{\displaystyle\int _{0}^{w}}\widetilde{\rho }\left (w^{ \prime }\right )\frac{\left (w -w^{ \prime }\right )^{1 -D\left (w\right )} +\left (w +w^{ \prime }\right )^{1 -D\left (w\right )}}{w^{ \prime 1 -D\left (w\right )}}dw^{ \prime }\widehat{\mathbf{w}} \nonumber  \\
\mathbf{g}_{outer}\left (w\right ) = +\frac{2\pi G}{l_{0}^{2}}{\displaystyle\int _{w}^{\infty }}\widetilde{\rho }\left (w^{ \prime }\right )\frac{\left (w^{ \prime } -w\right )^{1 -D\left (w\right )} -\left (w +w^{ \prime }\right )^{1 -D\left (w\right )}}{w^{ \prime 1 -D\left (w\right )}}dw^{ \prime }\widehat{\mathbf{w}} \nonumber \end{gather}
since, in general, the second integral in Eq. (\ref{eq7.4}) is not equal to zero.

In the special case $D =1$, the contribution from the outer shells is identically zero, since $\mathbf{g}_{outer}\left (w\right ) =0$ for this particular value of the dimension. The overall field is due just to the inner shells and, from Eq. (\ref{eq7.4}),  we have
$\mathbf{g}(w) =\mathbf{g}_{inner}\left (w\right ) = -\frac{4\pi G}{l_{0}^{2}}\int _{0}^{w}\widetilde{\rho }(w^{ \prime })dw^{ \prime }\widehat{\mathbf{w}}$, which corresponds to the expression in Eq. (\ref{eq7.3}) for $D =1$. Therefore, we conclude that our general result in Eq. (\ref{eq7.3}) is indeed valid for $1 \leq D \leq 3$.

\bibliographystyle{apsrev4-2}
\bibliography{mainNotes}

\end{document}